\documentclass[preprint,12pt]{elsarticle}

\usepackage{amssymb}
\usepackage{amsmath}
\usepackage{multirow}
\usepackage{adjustbox}
\usepackage{color}
\usepackage{float}
\usepackage{hhline}
\usepackage{graphicx}
\usepackage{subcaption}
\usepackage{xcolor}
\usepackage{makecell}
\usepackage{comment}
\usepackage{tikz}

\DeclareRobustCommand{\cnum}[1]{%
  \tikz[baseline=(char.base)]{
    \node[draw,circle,inner sep=1pt] (char) {#1};
  }%
}

\journal{Microprocessors and Microsystems}

\begin{document}

\begin{frontmatter}

\title{High-Performance NTT Accelerators for PQC leveraging Unified Redundant Arithmetic and Fine-Tuned Microarchitecture} 

\author[duth]{George Alexakis} 
\author[nokia]{Dimitrios Schoinianakis} 
\author[duth]{Giorgos Dimitrakopoulos} 

\affiliation[duth]{
    organization={Electrical and Computer Engineering, Democritus University of Thrace},
    city={Xanthi},
    country={Greece}
}

\affiliation[nokia]{
    organization={Nokia Bell Labs},
    city={Athens},
    country={Greece}
}

\begin{abstract}
Post-quantum cryptography and privacy-preserving technologies are expected to play a central role in future secure communication systems. Lattice-based PQC schemes such as ML-KEM (CRYSTALS-Kyber) and ML-DSA (CRYSTALS-Dilithium) rely heavily on large-degree polynomial arithmetic, making the Number Theoretic Transform (NTT) a key computational primitive. Although existing hardware accelerators exploit parallelism and pipelining to support both NTT and INTT, their efficiency is often limited by the overhead of modular reduction and correction steps, inverse-transform scaling operations, and suboptimal FPGA implementations. This work addresses these limitations by proposing parallel iterative NTT/INTT accelerators based on optimized unified butterfly units. We introduce a novel redundant number representation that eliminates conditional corrections for both Montgomery modulo multiplication and combined subtract–multiply operations, and integrate inverse-transform scaling into existing arithmetic hardware to avoid dedicated scaling units. Furthermore, we design hierarchical Montgomery multipliers that map efficiently onto FPGA DSP resources, reducing hardware cost while enabling high operating frequencies. FPGA-based experimental results demonstrate higher clock frequencies, reduced execution times, and competitive resource utilization, supporting efficient NTT acceleration for PQC and related privacy-preserving applications.
\end{abstract}

\begin{graphicalabstract}
\includegraphics[width=\columnwidth]{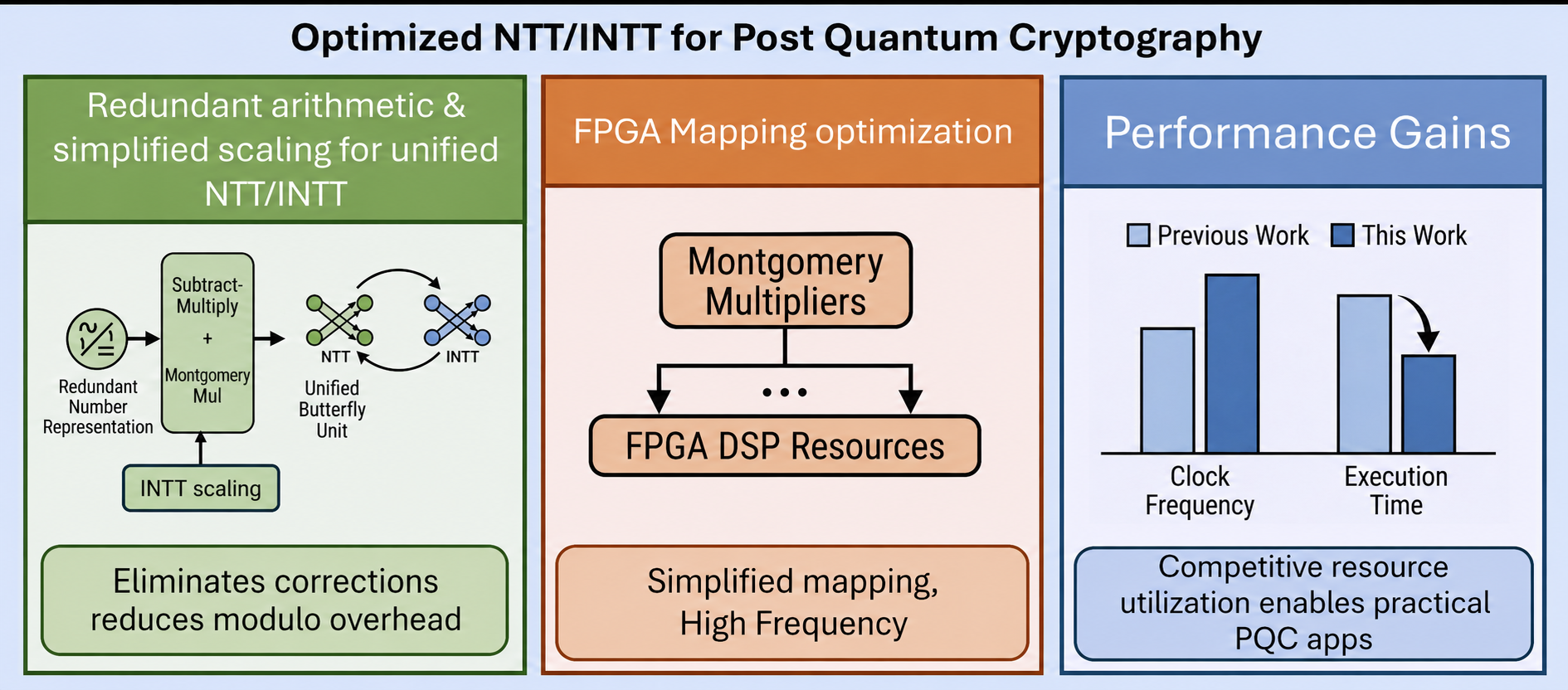}
\end{graphicalabstract}

\begin{highlights}
\item {\bf Unified redundant arithmetic for NTT/INTT:} Introduces a novel redundant number representation that extends Montgomery-based redundancy to support combined subtract–multiply operations, enabling a single butterfly unit to operate efficiently in both NTT and INTT modes while significantly reducing modulo correction overhead.

\item {\bf Microarchitectural Fine Tuning :} Merges INTT scaling operations into existing arithmetic hardware and hierarchically maps Montgomery multipliers onto FPGA DSP blocks, simplifying pipelining, lowering hardware cost, and enabling higher operating frequencies.

\item {\bf Consistent performance gains across PQC-relevant scales:} The proposed design achieves significant latency reductions over state-of-the-art NTT accelerators for runtime-programmable moduli, while maintaining comparable or smaller resource utilization and remaining competitive with fixed-modulus designs.
\end{highlights}


\end{frontmatter}

\section{Introduction}
\label{s:intro}
Post-Quantum Cryptography (PQC) and Privacy-Preserving Technologies (PPT) stand as the modern pillars of long-term secure communications, integrity, and data monetization. 
Their integration is particularly vital for the upcoming 6G era, where PQC is set to be a native security requirement from the beginning to defend against the threats posed by quantum computers \cite{nokia_sec_6G}. 
Future quantum computers will easily break the public-key algorithms that currently protect global digital infrastructure and rely on conventionally hard mathematical problems, such as integer factorization and the discrete logarithm, thereby creating an immediate security vulnerability~\cite{shor1994algorithms}.
At the same time, 6G networks are expected to leverage advanced PPTs, most notably Fully Homomorphic Encryption (FHE), to unlock secure data monetization and drive the next generation of privacy-centric Machine Learning (ML) and AI applications \cite{nokia_sec_6G}.

The Number Theoretic Transform (NTT) is a fundamental building block for PQC, as it enables efficient polynomial arithmetic that lies at the core of lattice-based schemes~\cite{liang2022number, fouque2018falcon, Hoffstein1998ntru}. Standardized PQC algorithms such as ML-KEM (formerly known as CRYSTALS-Kyber)~\cite{avanzi2019crystals} and ML-DSA (formerly known as CRYSTALS-Dilithium)~\cite{ducas2018crystals} rely on large-degree polynomial multiplications under modular arithmetic, and the use of NTT reduces their computational complexity from quadratic to near-linear. In many PQC schemes, polynomial multiplication is a well-known performance bottleneck~\cite{mert2022an}, making efficient NTT acceleration essential to achieve practical runtimes, with hardware acceleration offering substantial execution-time reductions compared to software-only implementations. Dedicated hardware can exploit extensive parallelism and apply domain-specific optimizations, such as optimized memory access patterns, reduced data movement, and highly efficient modular arithmetic implementations. As a result, such accelerators can deliver orders-of-magnitude improvements in both execution speed and energy efficiency.

State-of-the-art NTT hardware accelerators are typically organized around highly parallel~\cite{mu2023scalable, kemal2022cohantt} and deeply pipelined architectures~\cite{nguyen2024high, hirner2024proteus} that exploit the regular stage-wise structure of the transform. 
Parallel architectures consist of arrays of butterfly processing elements that perform modular additions, subtractions, and multiplications by precomputed twiddle factors, supported by dedicated modular reduction units. Memory is usually divided into multiple banks to enable parallel access without conflicts and overlapping computation and data movement~\cite{mu2023scalable, liu2024an}. Control logic orchestrates the sequencing of NTT stages, twiddle-factor addressing, and data permutations, while careful scheduling and interconnect design minimize latency and maximize throughput, allowing the accelerator to sustain high performance and energy efficiency across large polynomial sizes.

Implementing modular arithmetic in NTT accelerators, together with typical arithmetic operators, requires modular reduction steps that typically follow Barrett \cite{kong2010οptimizing} or Montgomery \cite{montgomery} reduction techniques. The hardware implementation of modular reduction represents a significant source of overhead, as it requires additional arithmetic operations and conditional checks beyond the core butterfly computations. This overhead increases hardware complexity and may also increase latency when targeting a specific clock frequency. Redundant data representations \cite{david2014faster, zewen2024a} have been proposed to reduce the cost of modular reduction by eliminating or postponing, as much as possible, the expensive modulo corrections.

In this work, we take a step further by introducing a new redundant representation that extends the redundancy used to eliminate conditional corrections in Montgomery multipliers to also support combined subtract–multiply operations. As a result, the same redundant representation can be applied to butterfly units operating in both NTT and INTT modes, further mitigating correction overhead without compromising correctness.

In addition to the redundant representation that reduces the cost of conditional corrections, we derive two additional microarchitectural improvements for the same two-mode butterfly units. First, the scaling (division) operations inherent to INTT are merged with existing arithmetic hardware, thereby eliminating the need for dedicated scaling units. Second, the Montgomery multipliers in the proposed design are hierarchically constructed from smaller multiplier blocks that naturally map onto the FPGA DSP block structure. This organization simplifies pipelining, reduces hardware cost, and enables high operating frequencies.

By combining a redundant data representation with targeted microarchitectural optimizations, the proposed design delivers substantial efficiency gains over state-of-the-art FPGA-based NTT accelerators. Notably, the architecture achieves higher operating frequencies without increasing pipeline depth, resulting in a direct reduction of absolute execution latency. Specifically, across the evaluated NTT sizes, the proposed accelerator reduces execution time by 35\% to 73\% compared to state-of-the-art accelerators implemented on the same FPGA platform with varying degrees of parallelism. These gains are sustained for arbitrary, runtime-programmable moduli, including those essential for PQC. Furthermore, the proposed approach maintains a similar or smaller hardware footprint compared to other runtime-configurable accelerators. Even when benchmarked against fixed-modulus accelerators, our solution achieves these latency reductions with only a marginal increase in hardware overhead.
\section{Background and related work}
\label{s:back}

The NTT transforms a sequence of $n$ coefficients of a polynomial into $n$ evaluations of that polynomial at specific points within the finite field. Therefore, multiplying two polynomials in this ``evaluation domain'' transforms the convolution-like operation of direct polynomial multiplication to an element-wise multiplication \cite{cohen}. An inverse NTT (INTT) then transforms the result back to the coefficient domain, yielding the coefficients of the product polynomial. 

Given a sequence of $n$ integers $a = (a_0, a_1, ..., a_{n-1})$ and a primitive $n$-th root of unity $\omega$ modulo a prime $q$ (meaning $\omega^n \equiv 1 \pmod{q}$ and $\omega^k \not\equiv 1 \pmod{q}$ for $1 \le k < n$), the forward NTT $A = (A_0, A_1, ..., A_{n-1})$ is defined as:
\begin{equation*}
A_k = \sum_{j=0}^{n-1} a_j \omega^{jk} \pmod{q} \quad \text{for } k = 0, 1, ..., n-1
\end{equation*}
The INTT recovers the original sequence:
\begin{equation*}
a_j = n^{-1} \sum_{k=0}^{n-1} A_k \omega^{-jk} \pmod{q} \quad \text{for } j = 0, 1, ..., n-1
\end{equation*}
Here, $n^{-1}$ is the modular multiplicative inverse of $n$ modulo $q$, i.e., $n\cdot n^{-1} \equiv 1 \mod q$.
Inputs $a_i$ are $W$-bits wide where $W = \lceil\log_2(q)\rceil$.

The direct computation of the NTT and INTT has quadratic complexity. However, similar to the Fast Fourier Transform (FFT), the NTT can be computed with logarithmic complexity using a divide-and-conquer approach~\cite{kaya2023post}. In this case, NTT of size $n$ is recursively divided into two NTTs of size $n/2$. The results are then combined using ``twiddle factors'', which are powers of the primitive $n$-th root of unity $\omega$. This recursive divide-in-half computation pattern concludes in a logarithmic number of steps and is computed using the well known FFT butterfly structures~\cite{heckbert1995fourier}. 

\begin{figure}[t]
\centering
\includegraphics[width=0.8\columnwidth]{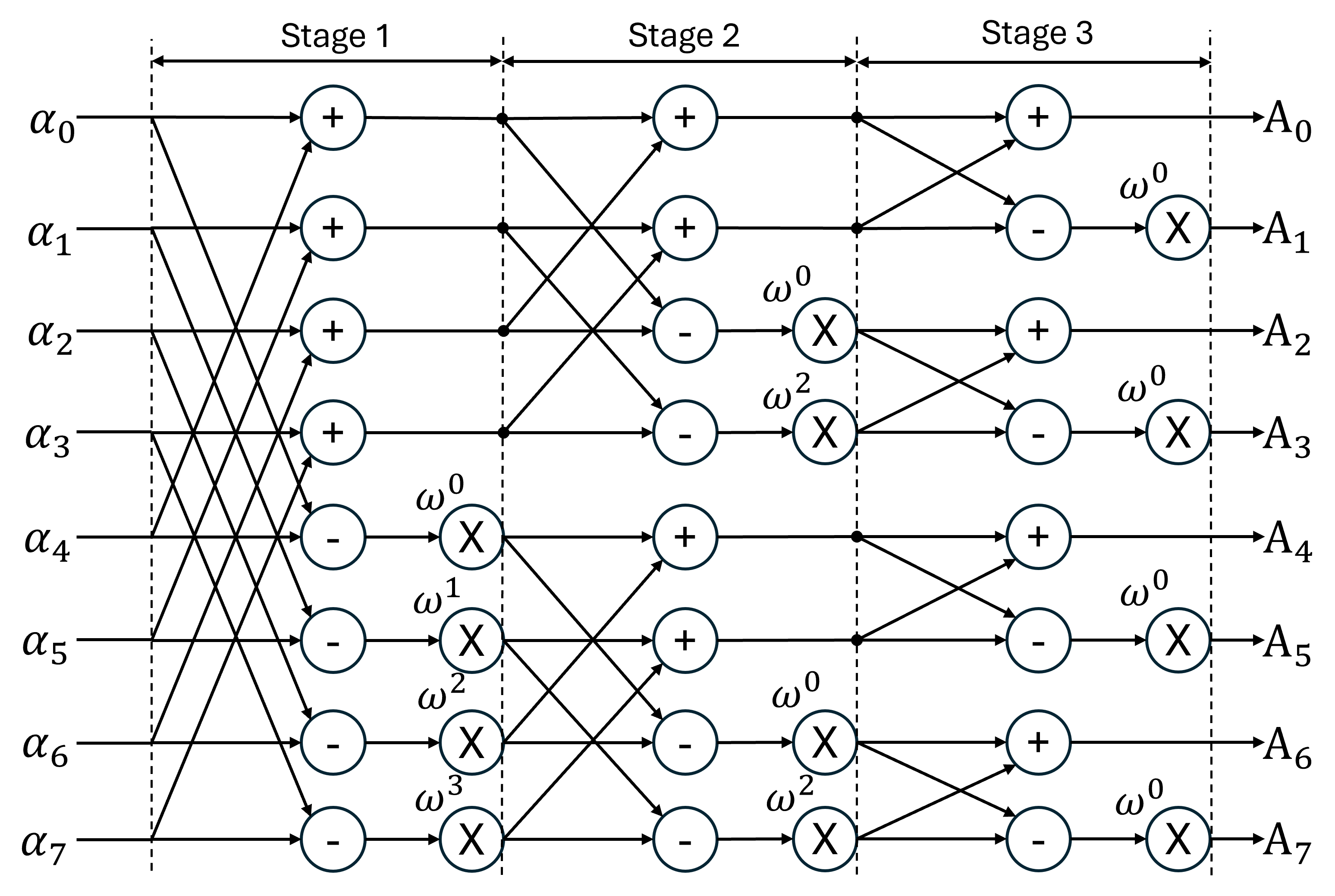}
\caption{Fast NTT computation following the Fast-Fourier Transform recursive paradigm.}
\label{fig:nttflow}
\end{figure}

Consider two adjacent elements in an intermediate stage of the NTT computation, say $A_j$ and $A_{j+n/2}$ (where $n$ is the current size of the sub-transform). A radix-2 butterfly operation computes the next stage values $A'_j$ and $A'_{j+n/2}$ as follows:
\begin{equation*}
\begin{aligned}
A'_j &= A_j + \omega^k \cdot A_{j+n/2} \pmod{q} \\
A'_{j+n/2} &= A_j - \omega^k \cdot A_{j+n/2} \pmod{q}
\end{aligned}
\end{equation*}
Here, $\omega^k$ is the twiddle factor, where the exponent $k$ depends on the stage of the algorithm and the index $j$. The twiddle factors used in each stage are specific powers of the primitive $n$-th root of unity, carefully chosen to ensure the correctness of the transform. The entire NTT dataflow for an 8-point using butterfly operations is shown in Fig.~\ref{fig:nttflow}. 

In essence, the NTT butterfly structure is the basic arithmetic unit that efficiently combines the results of smaller NTT computations, driven by the properties of the primitive roots of unity in the finite field. While radix-2 is the most common, other radices (like radix-4, radix-8) can also be used in NTT algorithms. These higher-radix butterflies process more than two inputs and outputs at a time, thus reducing the number of stages and potentially improving performance. 

NTT accelerators are designed using various approaches that can be categorized roughly in two basic categories: parallel and pipelined. Hierarchical or other hybrid approaches combine the main properties of these two fundamental approaches.

\subsection{Parallel Iterative NTT hardware accelerators}

\begin{figure}[t]
\centering
\includegraphics[width=0.65\columnwidth]{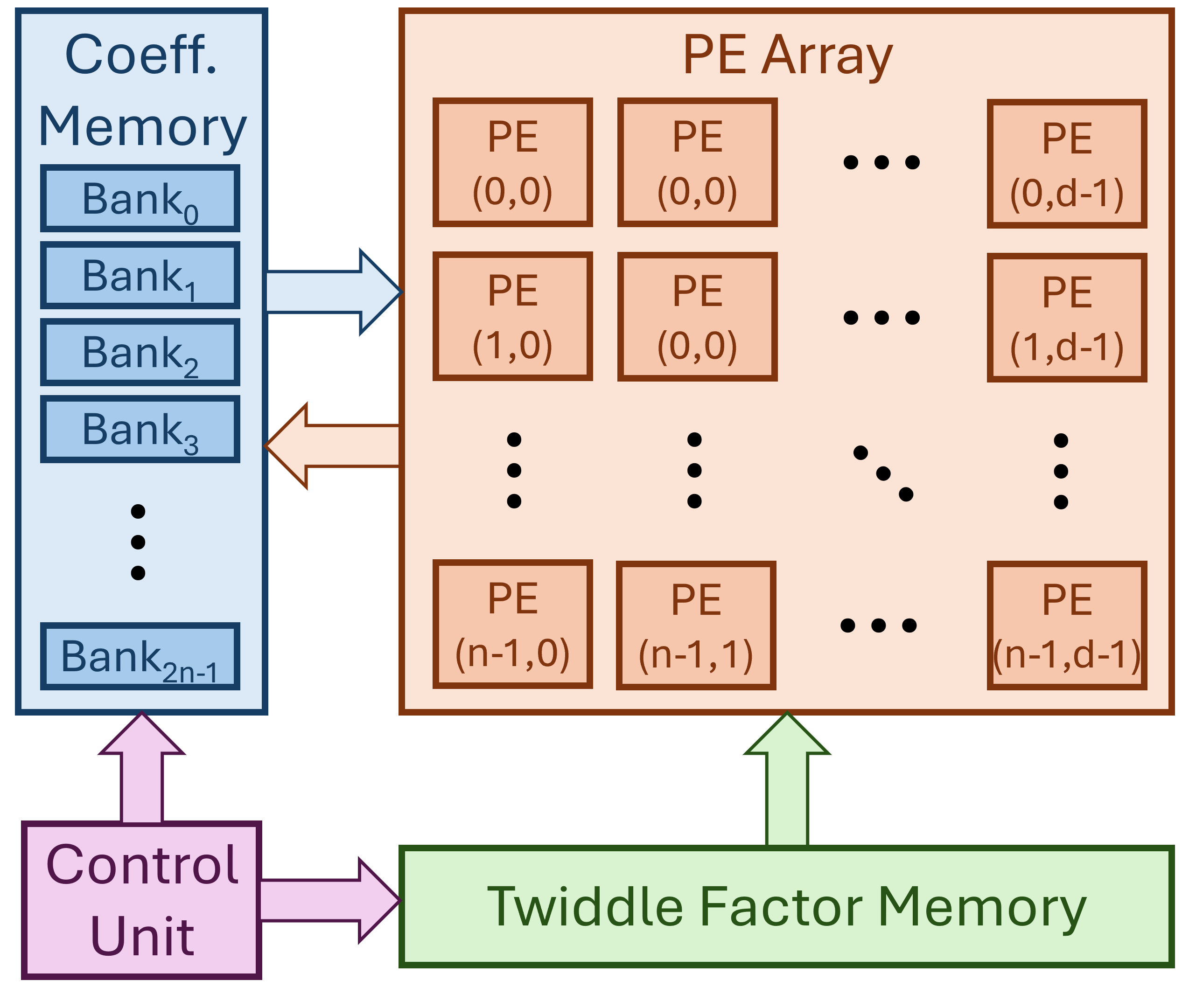}
\caption{Parallel NTT accelerator that computes NTT iteratively using an array of PEs (effectively butterfly units) and multiple parallel memory banks.}
\label{fig:iterative}
\end{figure}

Parallel NTT accelerators typically consist of an array of butterfly units, also referred to as processing elements (PEs), and multi-banked memories. These components are supported by an address generation logic that computes NTTs of arbitrary sizes iteratively. A typical organization of a parallel NTT accelerator is shown in Fig.~\ref{fig:iterative}. The design exploits parallelism in both computation, through multiple butterfly units, and memory access, through parallel memory banks. 

Typical representatives of such parallel NTT accelerators are the designs presented in~\cite{di2023vlsi} and~\cite{kemal2022cohantt}, while~\cite{mert2022an} provides a comprehensive study of flexible design methodologies, evaluating hand-crafted parametric generators, as well as other implementations. These designs employ one or multiple butterfly units in parallel and differ in how they perform modular multiplication. 

Other approaches use a two-dimensional array of butterfly units. These accelerators usually provide the highest performance, but they also have high complexity. The main challenge comes from memory conflicts caused by the very high bandwidth requirements. To address this issue, the design in \cite{liu2024an} adopts a different NTT algorithm, called CG-NTT, which preserves the same memory access pattern across all stages. This property allows coefficients to be easily distributed across memory banks for efficient access. A drawback of this approach is that the NTT can no longer be computed in place, which doubles the memory requirements. However, the method proposed in this work reduces this overhead by half.

In contrast, CFNTT \cite{chen2021cfntt} targets the standard NTT algorithm. It represents NTT coefficient indices using a mixed-radix format, enabling each coefficient to be mapped to a specific memory bank and address through a tailored memory mapping scheme. This approach supports in-place NTT computation and avoids the additional memory overhead introduced by CG-NTT.

A similar but more general approach is presented in \cite{mu2023scalable}. This work relies on a unified radix representation of indices and an efficient index generation mechanism for memory addressing. In addition, it introduces a read-after-write conflict-free access scheme that eliminates pipeline idle cycles when certain conditions on the NTT size and butterfly array latency are satisfied. As a result, the design achieves near-ideal execution cycles for specific butterfly array sizes.

OpenNTT~\cite{krieger2025openntt} is an NTT generator that adopts similar principles for the automatic design efficient parallel NTT accelerators.

\subsection{Pipelined NTT Hardware Accelerators}

In contrast to generic iterative architectures, pipelined architectures follow a linear pipelined structure that targets the removal of irregular memory accesses by appropriately delaying data elements (inputs or intermediate results) before being used in subsequent stages of the NTT similar to high-performance pipelined FFT accelerators~\cite{garrido2022survey}. 

\begin{figure}[t]
  \centering
  \begin{subfigure}{\textwidth}
    \centering
    \includegraphics[width=\linewidth]{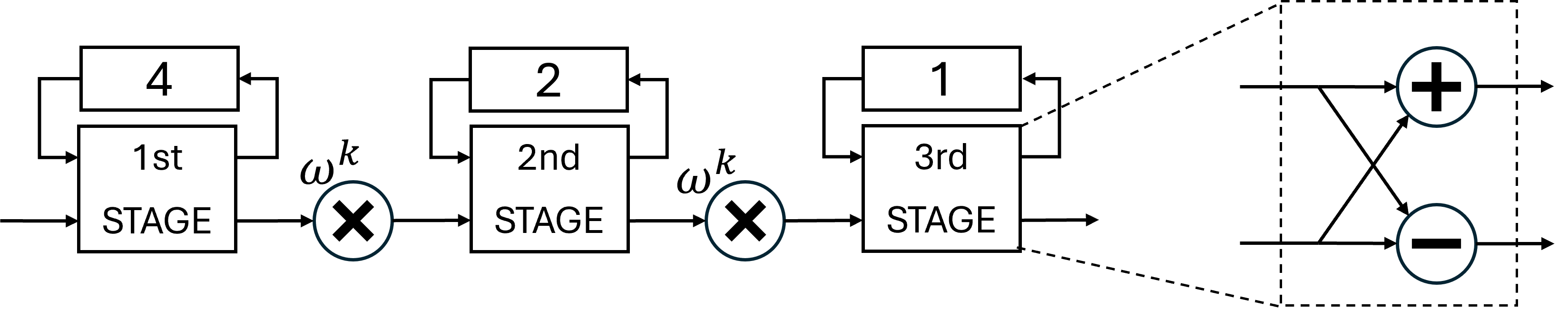}
    \caption{Single Data Flow Pipelined NTT for $N=8$}
    \label{fig:sdfarch}
  \end{subfigure}
  \hfill
  \begin{subfigure}{\textwidth}
    \centering
    \includegraphics[width=\linewidth]{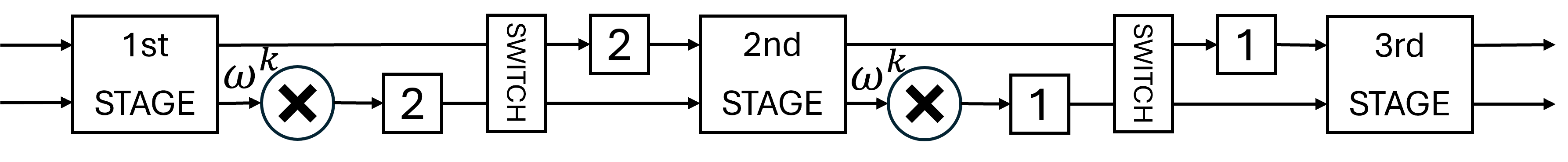}
    \caption{Pipelined NTT for $N=8$ following the Multi-path Delay Commutator architecture.}
    \label{fig:mdcarch}
  \end{subfigure}
  \caption{The two main forms of pipelined NTT hardware accelerators. Each numbered rectangle denotes a delay buffer, where the number inside indicates the number of clock cycles by which the corresponding input is delayed. }
  \label{fig:dataflowarch}
\end{figure}

Pipelined architectures for the NTT are primarily categorized into Single-path Delay Feedback (SDF) and Multi-path Delay Commutator (MDC) designs, which differ in their data-shuffling mechanisms and hardware efficiency. Examples of the two representatives are illustrated in Fig.~\ref{fig:dataflowarch}. SDF architectures utilize feedback loops with delay registers to store intermediate butterfly outputs; however, they typically suffer from lower hardware utilization as processing elements remain idle during feedback cycles. In contrast, MDC architectures employ a feed-forward structure with commutators and delay elements to split the input into multiple parallel streams, achieving higher throughput and potentially full processing-element utilization at the cost of higher memory overhead. SDF is often chosen for area-constrained designs due to its small memory footprint, while MDC is better suited for high-performance cryptographic accelerators. 

The work in \cite{nguyen2024high} presents a high-speed accelerator for PQC, specifically for ML-KEM and ML-DSA. The radix-2 and radix-4 MDC architectures are proposed and the modular multiplication algorithm used is K-RED, an algorithm that is optimized for modular multiplication by constants.

Proteus~\cite{hirner2024proteus} is a configurable, design-time hardware generation framework that produces synthesizable pipelined NTT architectures based on Radix-2 SDF and MDC structures for user-defined parameters. It includes parameterized designs for fundamental arithmetic components, including integer multipliers, modular reduction units, and compact butterfly modules that all employ word-level Montgomery reduction.

For NTTs that require large-integer arithmetic, as used in PPT such as Zero-Knowledge Proofs (ZKP) and FHE, our previous work in \cite{alexakis2025high} introduced a multi-path delay feedback design. This approach functions as a large-word SDF architecture by splitting computation across multiple paths, each operating on smaller, uniform word sizes, improving scalability and achievable clock frequency.

\subsection{Other approaches for NTT hardware acceleration}

Hierarchical approaches, while very common in software implementation, are less popular on hardware accelerators. These approaches decompose a large NTT into smaller size NTTs and then compute them using one of the previously discussed methods. For instance, PipeZK~\cite{zhang2021pipezk} combines hierarchical and pipelined approaches. For the decomposition, the 4-step NTT method is utilized, and for the smaller NTT computations, an SDF architecture is used. The accelerator is optimized for ZKPs, thus it is one of the few that operate on very large numbers and NTT sizes. SAM~\cite{wang2023sam} generalizes decomposition to multiple dimensions adopting the decomposition strategy to the size of the NTT required by the selected application domain. 

NTTGen \cite{Yang2022nttgen} is an FPGA-oriented framework that automatically generates low-latency NTT accelerators for FHE. It allows designers to specify high-level parameters, such as polynomial size and degree of parallelism, while hiding low-level hardware implementation details. To efficiently manage the irregular data access patterns of NTT, NTTGen uses a streaming permutation network, avoiding costly crossbars and complex memory layouts. Similarly, AutoNTT \cite{kumarathunga2025autontt} is an automated framework for designing FPGA-based NTT accelerators. It supports a wide range of NTT configurations by adjusting parameters such as transform size and parallelism, enabling the generation of application-specific optimized pipelines. AutoNTT also supports parallel, pipelined, and hybrid architectures. To reduce the cost of NTT implementation in FPGAs, in~\cite{ni2023towards} block memories are replaced by appropriately scheduled FIFO buffers.

In contrast to other approaches, MeNTT~\cite{li2022mentt} targets the in-memory computation of NTT. They propose bit serial algorithms to compute modular addition, subtraction and multiplication in memory and a memory mapping strategy that reduces routing overhead and optimizes the NTT dataflow. 


\subsection{Low-level Microarchitectural optimizations}

Many studies focus on low-level optimizations of fundamental modular operations, such as addition, subtraction, and multiplication. One common approach, also adopted in this work, is the use of redundant representations to streamline modulo computation.

Redundant Montgomery representations were first systematically explored by Pu et al.~\cite{pu2009montgomery}, who showed that allowing operands and intermediate results to reside in an extended dynamic range eliminates the need for final conditional subtractions in Montgomery exponentiation. This work establishes the baseline redundancy model for Montgomery multipliers and serves as a foundation for our design that trades datapath width for reduced control and correction overhead.

Similarly, \cite{david2014faster} proposes a redundant representation designed to minimize corrections within butterfly operations using Shoup’s modular multiplication. By allowing two of the three butterfly inputs to exceed the modulus, the authors demonstrate that the outputs remain bounded within this expanded range, allowing successive butterfly operations on redundant data. Targeting vector processors, the work of \cite{zewen2024a} introduces ``lazy reduction'' techniques that leverage redundant representations to skip corrections in certain signed Montgomery multiplications. While lazy reductions in custom NTT hardware accelerators do not eliminate the physical circuitry required for reductions, they can potentially lower power consumption by enabling the reduction logic only when strictly necessary.

Other approaches, such as HRCIM-NTT~\cite{Zhang2025hrcimntt}, employ hybrid data types to speedup modular multiplication: the twiddle factors are represented in 2's complement, and the polynomial coefficients are converted into the minimally redundant radix-4 representation leading to a radix-4 Booth algorithm.

KNightCore~\cite{taghavi2025knightcore} and LightNTT~\cite{design2026lightntt} propose unified datapaths for NTT, INTT, and point-wise multiplication. Both architectures reuse the ROM storing the NTT twiddle factors and the NTT butterfly hardware to also perform point-wise multiplications. Additionally, they employ a hardware-friendly Barrett modular multiplication algorithm optimized using shift-and-add operations, thereby completely avoiding the use of DSP blocks in FPGA implementations.

The work of~\cite{bisheh2021high} proposes a high-speed NTT-based polynomial multiplication accelerator focusing on efficient modular reduction within the butterfly units. Specifically, the architecture further introduces  a hardware-friendly modular reduction algorithm, which requires few resources without additional cost of memory utilization.

Finally, the work in~\cite{guo2023highly} targets the simplification of scaling (division) operations required in INTT by merging scaling unit with modular addition. A similar merging approach is proposed in this work. However, the merging method differs in the order of operations performed that fits better to the unified butterfly unit proposed in this work.

\section{Proposed architecture}
\label{s:prop}

\begin{figure}[t]
\centering
\includegraphics[width=0.6\columnwidth]{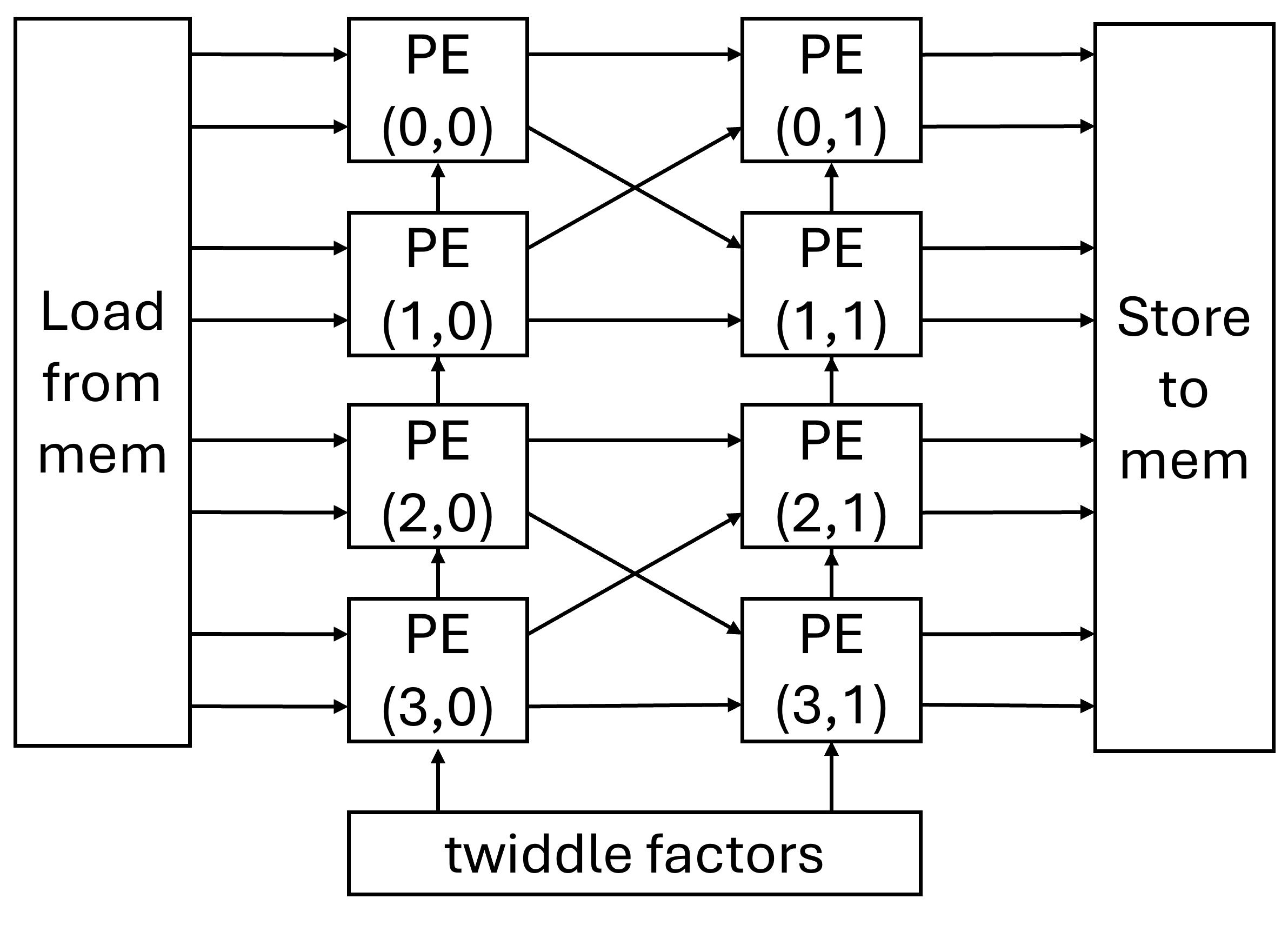}
\caption{Internal connectivity of a $4\times 2$ PE array.}
\label{fig:pecon}
\end{figure}

The proposed approach is based on the parallel NTT accelerator architecture shown in Fig.~\ref{fig:iterative}. The processing elements (PEs) are arranged in a 2D grid, where each PE contains a butterfly unit. Each butterfly takes two input elements and a twiddle factor and produces the outputs for the next NTT stage. Communication between PEs follows a butterfly-style interconnect: PEs within the same column are not connected, and data flows only from left to right. This data movement is illustrated in Fig.~\ref{fig:pecon} for a $4 \times 2$ array. The architecture reuses the same PE array to compute the full NTT. In general, the array can be organized as a $w \times d$ grid of PEs, where $d \le \log(w) + 1$.

To address complex memory access patterns without increasing storage complexity or modifying the NTT algorithm, as in \cite{liu2024an}, we adopt the conflict-free memory mapping scheme proposed in \cite{mu2023scalable}. Under this scheme, each coefficient is mapped to a unique memory bank and a unique address within that bank.

Following the same architectural approach, pipeline bubbles in the dataflow can be completely avoided as long as the latency of each PE is less than or equal to $\frac{N}{2\, w\, d\, 2^d}$, where $N$ is the NTT size, $w$ is the number of rows in the PE array, and $d$ is the number of columns. Although this constraint limits the number of feasible design configurations, we design low-latency PEs, as described in Subsection~\ref{ss:dsp}, that support most practical configurations.

\begin{figure}[t]
    \centering
    \begin{subfigure}{\textwidth}
        \centering
        \includegraphics[width=0.7\linewidth]{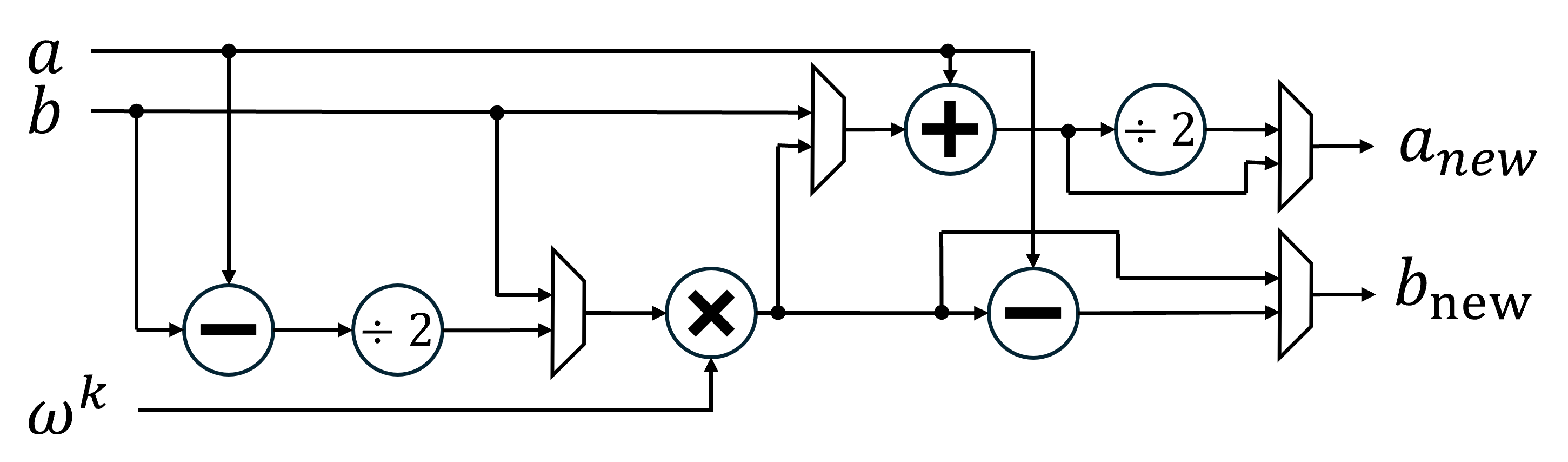}
        \caption{Unified Butterfly that can operate in NTT and INTT modes.}
        \label{fig:ctgs1}
    \end{subfigure}
    \begin{subfigure}{0.7\textwidth}
        \centering
        \includegraphics[width=\linewidth]{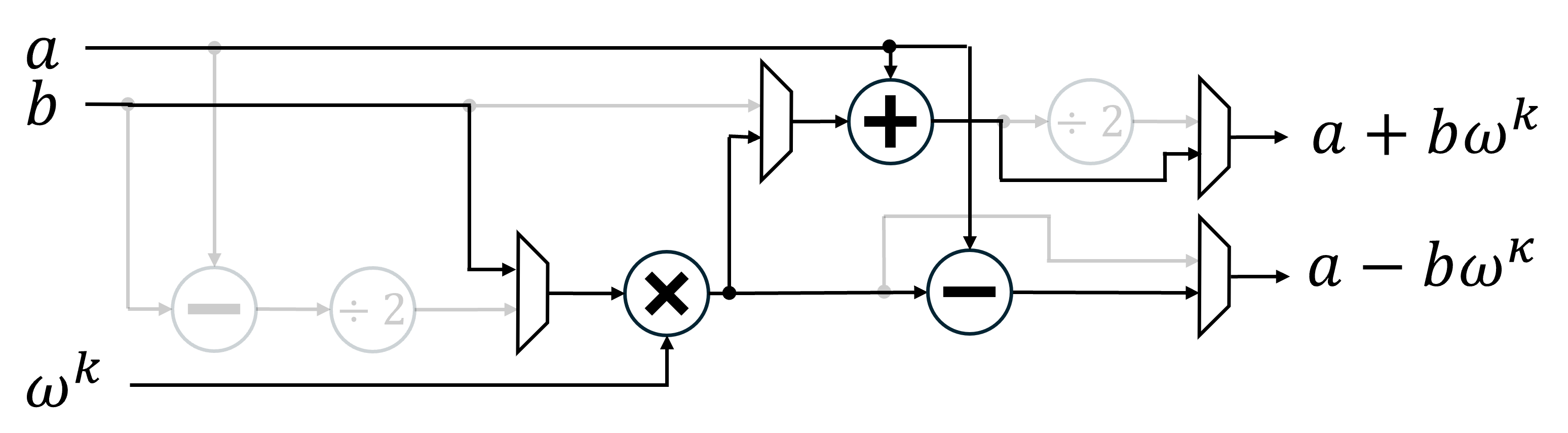}
        \caption{Data flow and computation in NTT mode}
        \label{fig:ct}
    \end{subfigure}
    \begin{subfigure}{0.7\textwidth}
        \centering
        \includegraphics[width=\linewidth]{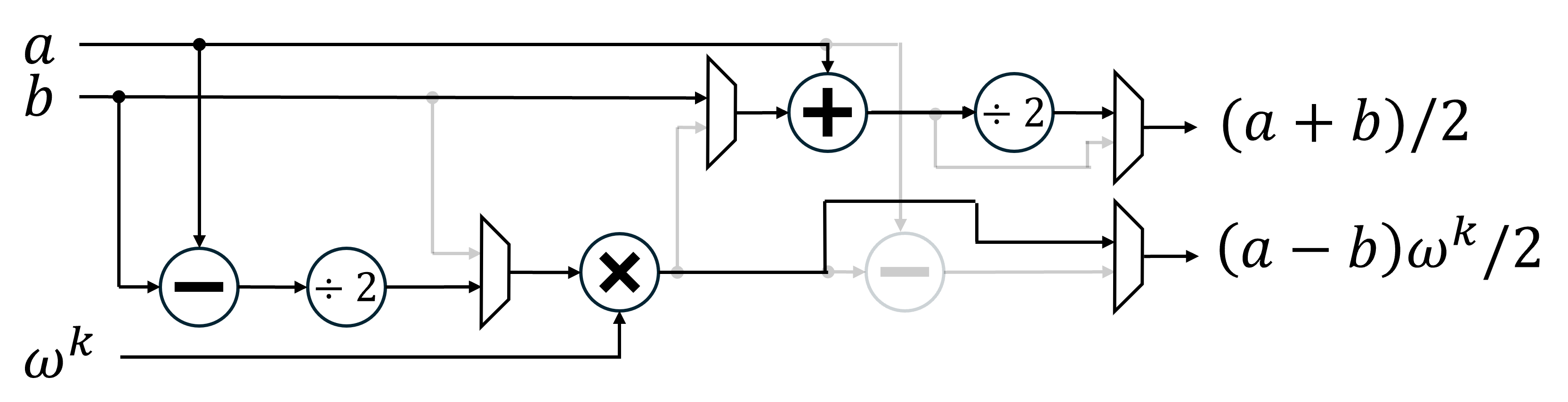}
        \caption{Data flow and computation in INTT mode}
        \label{fig:gs}
    \end{subfigure}
\caption{The unified butterfly structure from~\cite{zhang2020highly} serves as the PE for the proposed architecture. It is configurable for both NTT and INTT modes, with all arithmetic operations performed modulo $q$.}
\label{fig:ctgs}
\end{figure}

\subsection{Microarchitecture of each PE}

Each PE implements a butterfly unit that supports both NTT and INTT, a common feature in state-of-the-art accelerators. In the proposed design, we adopt and further optimize the unified butterfly architecture introduced in \cite{zhang2020highly} and shown in Fig.~\ref{fig:ctgs}. The PE operates in two modes, corresponding to NTT and INTT computation, and consists of two modular subtractors, one modular adder, one Montgomery multiplier, and two divide-by-two units used exclusively in INTT mode. To optimize resource utilization, the modular adder is shared between NTT and INTT operations. In addition, delay registers (not shown in Fig.~\ref{fig:ctgs}) are integrated to the inputs to match the pipeline depth of the parallel arithmetic units, ensuring synchronized data flow throughout the butterfly computation.

\begin{figure}[t]
\centering
\includegraphics[width=0.8\columnwidth]{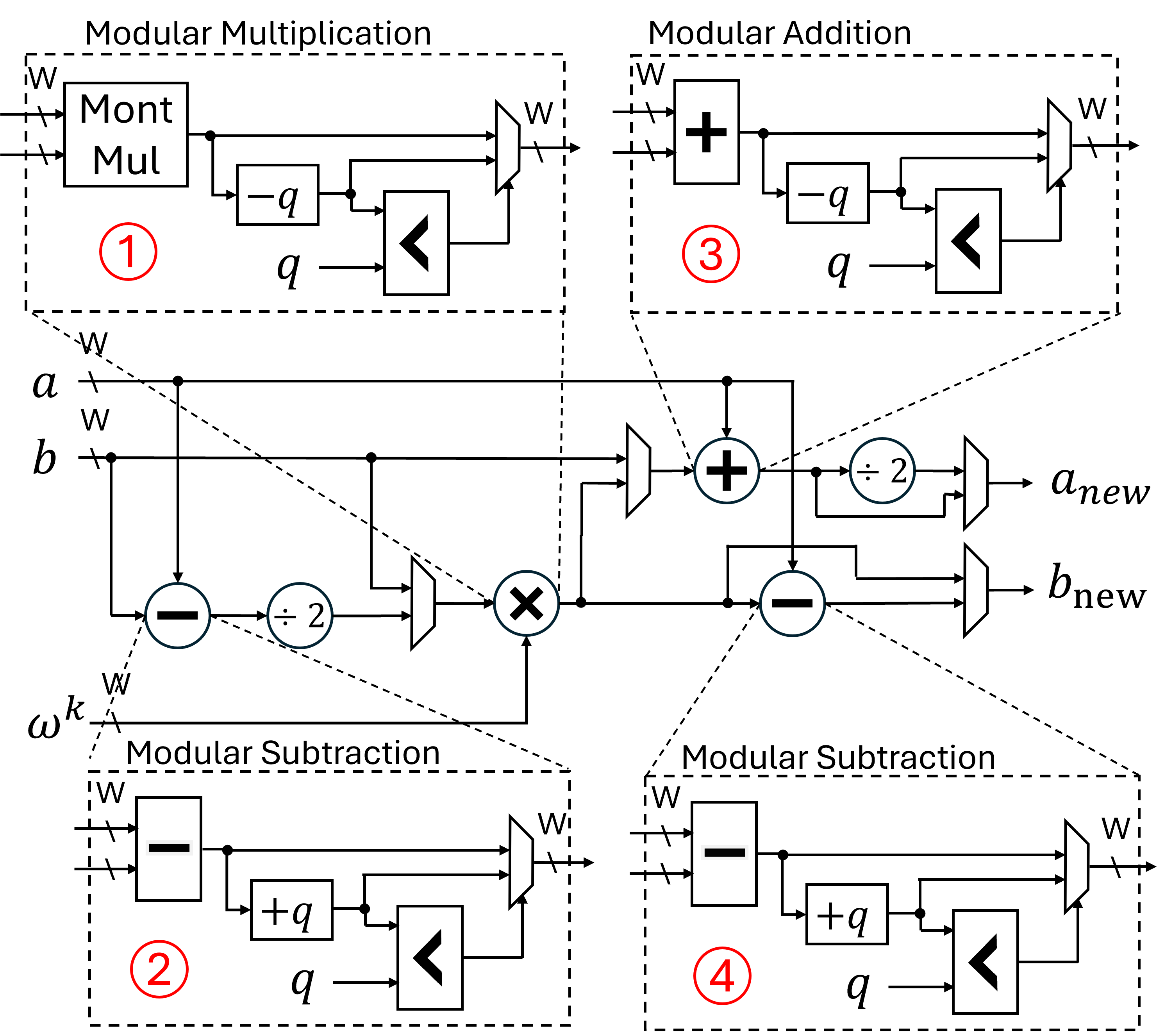}
\caption{The Unified butterfly unit structure proposed in \cite{zhang2020highly} highlighting the conditional correction steps needed after each modulo arithmetic operator. Multiplication follows the architecture proposed by Montgomery~\cite{montgomery}. Inputs, intermediate results and outputs are restricted to the canonical range \([0, q-1]\) thus represented with $W=\left\lceil\log q\right\rceil$ bits.}
\label{fig:pe}
\end{figure}

Data at the inputs and outputs of each modular arithmetic unit are assumed to lie within the non-redundant dynamic range $[0, q-1]$. To maintain this range, each modular arithmetic unit in Fig.~\ref{fig:ctgs} includes a conditional correction stage, as highlighted in Fig.~\ref{fig:pe}. This stage checks whether the result exceeds the modulus (q), as required for addition and Montgomery multiplication, or becomes negative in the case of subtraction, and applies the appropriate correction.

These conditional checks are performed sequentially after the main arithmetic operation. As a result, they increase the combinational delay, which can degrade clock frequency and in effect overall throughput or require the insertion of additional pipeline stages to compensate for the added delay.

The proposed design refines the PE microarchitecture shown in Fig.~\ref{fig:pe} to achieve three primary objectives: (a) the removal of modulo correction steps \cnum{1} and \cnum{2} through a new redundant data representation that covers both NTT and INTT operations; (b) the elimination of one scaling operation and the integration of the remaining one into modulo correction step \cnum{3}; and (c) the optimized mapping of the Montgomery multiplier to FPGA DSP blocks. The following sections provide a detailed description of how these goals are realized.

\subsection{Eliminating post-multiplication modulo corrections with redundant number representation}

The output of the mixed-mode unified butterfly shown in Fig.~\ref{fig:pe} computes the following depending on the selected mode:
\[ 
\begin{array}{ll} 
\text{NTT:} & a^{NTT}_{\text{new}} = a + b\omega^k \,\pmod{q} \\
            & b^{NTT}_{\text{new}} = a - b\omega^k \,\pmod{q} \\
\\ 
\text{INTT:} & a^{INTT}_{new} = (a + b)/2 \pmod{q} \\
             & b^{INTT}_{\text{new}} = (a - b)\omega^k/2 \pmod{q}
\end{array}
\]

\noindent The modular division by two for INTT does not increase the range of the output therefore it is omitted for brevity in the following output range analysis. How these scaling operations are handled in this work is described in detail in Section~\ref{ss:div2}.

In this work, all values are represented in Montgomery form, meaning that each value is implicitly multiplied by a constant $R$, known as the Montgomery radix. The radix \(R\) is chosen as a power of two and is larger than the modulus \(q\) \cite{montgomery}. For any value \(a \bmod q\), its Montgomery representation is defined as \(\overline{a} = aR \bmod q\).

In NTT mode, the Montgomery multiplier computes \(b\,\omega^k\). To eliminate in NTT mode the modulus correction step after Montgomery multiplication \cnum{1}, we adopt the approach presented in \cite{pu2009montgomery}. Instead of restricting operands to the canonical range \([0, q-1]\), a redundant representation in the range \([0, 2q-1]\) is permitted, where each residue class \(x \bmod q\) may be represented by either \(x\) or \(x+q\). Under this representation, and assuming that \(R > 4q\), in~\cite{pu2009montgomery} it was shown that the output of the Montgomery multiplication in NTT mode is guaranteed to lie within \([0, 2q-1]\).

Because the output remains within the same extended redundant dynamic range as the multiplier inputs, the correction step \cnum{1} can be omitted for the computation of \(b\,\omega^k\) in NTT mode.

However, in the unified butterfly unit shown in Fig.~\ref{fig:pe}, the same Montgomery multiplier is also used to compute the term \((a - b)\omega^k\) needed in INTT mode. Therefore, to eliminate the post-multiplication modulo correction step \cnum{1} in both NTT and INTT modes, it is necessary to \emph{extend the redundant number representation} used for NTT to \emph{a new form that supports uniformly the multiplication operations in both modes}. 

To identify the parameters of this new redundant form, we start by analyzing the output range of the multiplier in INTT mode assuming the same redundant input representation for inputs $a, b, \omega^k$. More specifically, the Montgomery form of inputs $a, b$, and $\omega^k$ is assumed to lie in the following dynamic range:
\begin{equation}
0 \leq \overline{a} <2q \qquad
0 \leq \overline{b}<2q \qquad
0 \leq \overline{\omega^k}<2q
\label{e:inp-bounds}
\end{equation}

To evaluate the dynamic range of \((a - b)\omega^k \pmod{q}\) computed in INTT mode, we first determine the dynamic range of the difference $a - b$. Its Montgomery representation is given by \(\overline{a - b} = \overline{a} - \overline{b}\), which yields according to~\eqref{e:inp-bounds}
\begin{equation}
-2q < \overline{a-b} < 2q
\label{e:tsubbounds}
\end{equation}

Similarly, for the sum $a + b$ computed also in INTT, we can compute according to~\eqref{e:inp-bounds} that $0 < \overline{a + b} < 4q$. To align the output range of both the addition and subtraction operations, we introduce an extra term $2q$ to subtraction, resulting in:
\begin{equation}
0 < \overline{a-b} + 2q < 4q
\label{e:inc-bound-sub}
\end{equation}
This adjustment ensures that the output range of $a - b + 2q$ matches that of $a + b$ exactly. Since all operations are performed modulo $q$, adding the $2q$ term only introduces redundancy at the output of the subtractor without affecting the final result. This increased value by $2q$ is passed later on to the multiplier to compute \((a - b + 2q)\omega^k \pmod{q}\).

The computation of $b^{\text{INTT}}_{\text{new}} = (a - b + 2q)\omega^k \pmod{q}$ can be rewritten using the Montgomery formula as follows:
\begin{align}
\overline{b^{\text{INTT}}_{\text{new}}}& =\overline{(a - b + 2q)\omega^k} \nonumber\\
& =\frac{\overline{(a - b + 2q)}\, \overline{\omega^k}+ 
q[\overline{(a - b + 2q)}\, \overline{\omega^k}\, \mu]_R}{R} \label{e:bnewmont}
\end{align}
where $\mu$ represents the Montgomery inverse of $q$ modulo $R$.

Since the term $[\overline{(a - b + 2q)}\, \overline{\omega^k}\, \mu]_R$ is a residue modulo $R$ its value is always less than $R$. When multiplied by $q$ as done in~\eqref{e:bnewmont} the following output range is obtained:
\begin{equation}
0 \le [\overline{(a - b + 2q)}\, \overline{\omega^k}\, \mu]_R{R} \le qR
\label{e:rest}
\end{equation}

Using bound~\eqref{e:inp-bounds} for $\omega^k$ and~\eqref{e:inc-bound-sub} for $a - b + 2q$ we can compute that 
\begin{equation}
0 \le \overline{(a - b + 2q)}\, \overline{\omega^k} \le 4q\, 2q = 8q^2
\label{e:sub-mul}
\end{equation}

\noindent Therefore, by summing inequalities~\eqref{e:rest} and~\eqref{e:sub-mul} we get that 
\begin{equation}
0 \le \overline{(a - b + 2q)}\, \overline{\omega^k}+ 
q[\overline{(a - b + 2q)}\, \overline{\omega^k}\, \mu]_R \le qR + 8q^2
\end{equation}
Dividing all sides by $R$ as needed to form $\overline{b^{\text{INTT}}_{\text{new}}}$ yields
\begin{equation}
0 \le \overline{b^{\text{INTT}}_{\text{new}}}\le q + 8q^2/R
\label{e:intt-range}
\end{equation}
By selecting $R > 8q$ transforms~\eqref{e:intt-range} to
\begin{equation}
0 \le \overline{b^{\text{INTT}}_{\text{new}}}\le 2q
\label{e:last}
\end{equation}

In this way, we have shown that the output range of the Montgomery multiplier, when computing $b^{\text{INTT}}_{\text{new}}$ in INTT mode with $R > 8q$, follows the same extended dynamic range as variables $a$, $b$, and $\omega^k$. Consequently, when the Montgomery multiplier completes the subtract-multiply operation $(a - b + 2q)\omega^k \pmod{q}$ using redundant inputs, there is no need for any additional modulo correction steps. Recall that, following~\cite{pu2009montgomery}, the same conclusion applies when the Montgomery multiplier computes $b\,\omega^k$ in NTT mode, assuming $R > 4q$. Since $R > 8q$ inherently satisfies the $R > 4q$ requirement, this choice enables the complete removal of step \cnum{1} in Fig.~\ref{fig:pe} in both NTT and INTT operations.

Furthermore, the proposed redundant data representation and its unified application to the subtract-multiply operation allows also for the removal of the correction step \cnum{2}. Although the subtractor computing $a - b + 2q$ produces an output in the dynamic range $[0, 4q-1]$ as shown in~\eqref{e:inc-bound-sub}, which technically exceeds the extended range of $[0, 2q-1]$, this result is consumed exclusively by the subsequent Montgomery multiplier. Because this intermediate value is never used independently elsewhere in the design, and since it has been proven in~\eqref{e:last} that a $[0, 4q-1]$ range of  $a - b + 2q$  does not compromise the final output range of the combined subtract-multiply operation, the intermediate modulo correction for the subtractor can be safely eliminated.

\begin{figure}[t]
\centering
\includegraphics[width=0.9\columnwidth]{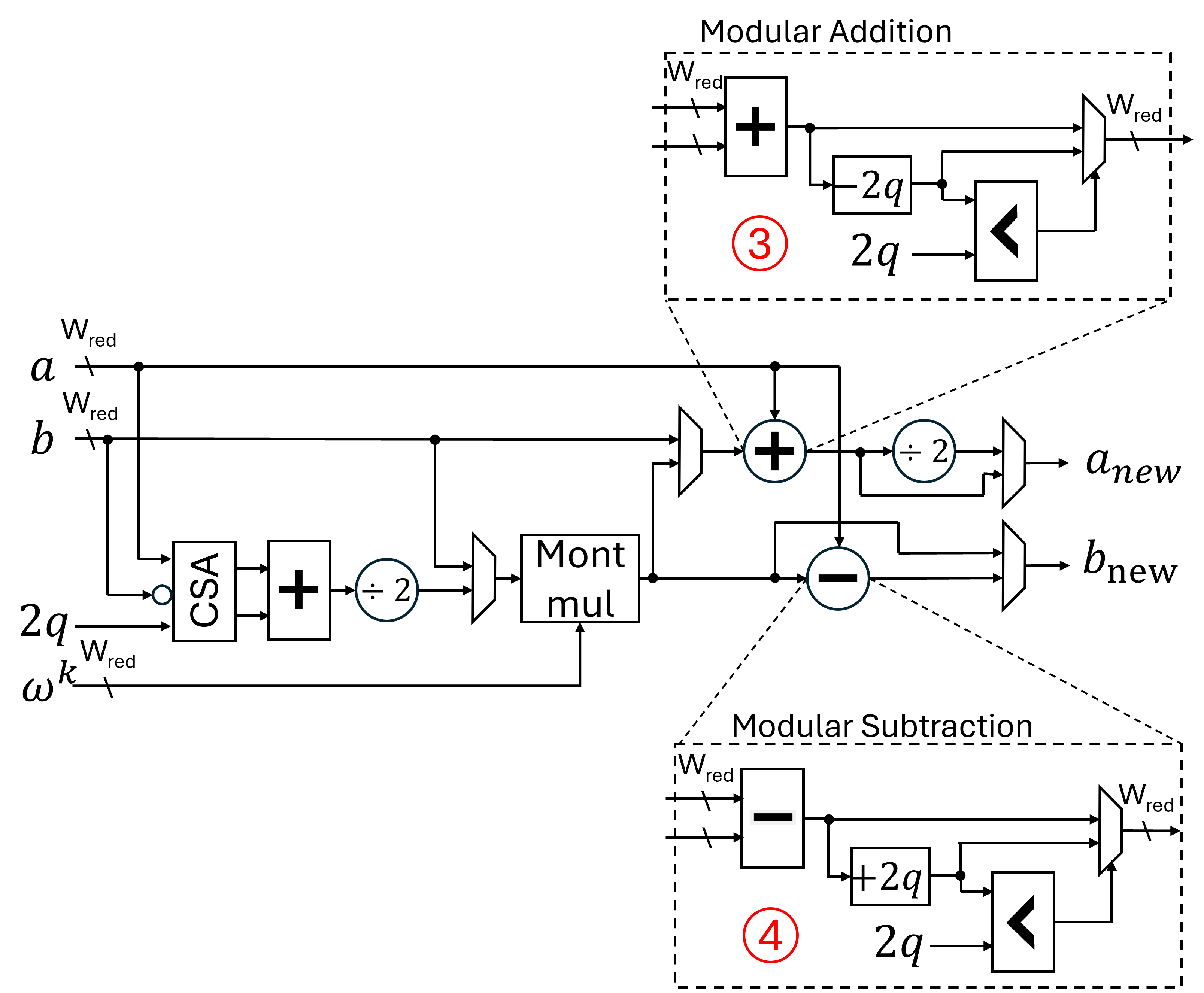}
\caption{The unified butterfly unit covering both NTT and INTT operating on an extended redundant dynamic range $[0, 2q-1]$. Since all inputs, outputs and intermediate results follow the same Montgomery form using a radix $R > 8q$ the bitwidth of each variable increases to $W_{\text{red}}=\left\lceil\log q\right\rceil+3$.}
\label{fig:pe_redundant}
\end{figure}

The resulting unified butterfly unit, which supports both NTT and INTT computations through our proposed redundant data representation, is illustrated in Fig.~\ref{fig:pe_redundant}. As previously detailed, the correction steps \cnum{1} and \cnum{2} from the original architecture in Fig.~\ref{fig:pe} have been eliminated. Conversely, correction steps \cnum{3} and \cnum{4} are retained, though their constants are updated from $q$ to $2q$. 
This adjustment ensures that the outputs of the final addition and subtraction stages in NTT mode follow the same dynamic range as their inputs and therefore remain within the extended dynamic range \([0, 2q-1]\), which is now applied uniformly throughout the datapath. Furthermore, unlike the original design, all input, output, and intermediate signals in the improved unit operate with a bit-width of $\lceil\log q\rceil+3$. These three additional bits are necessary to satisfy the requirement that the Montgomery radix $R$ is at least eight times larger than the modulus $q$. For this reason, $R$ is chosen as the next power of two exceeding $8q$.

\subsection{Eliminating and simplifying division-by-2 units needed in INTT mode}
\label{ss:div2}

The unified butterfly unit includes two "divide-by-2" scaling operations that are active only during INTT mode. The first one placed at the input of the Montgomery multiplier can be eliminated, by shifting its cost to offline precomputation. Instead of supplying the multiplier with standard INTT twiddle factors ($\omega^{-k} \pmod{q}$), we utilize twiddle factors that are already divided by two: $\frac{\omega^{-k}}{2} \pmod{q}$. Because these twiddle factors are constant for a specific modulus $q$, they can be calculated offline. Consequently, this transformation achieves the required scaling without incurring any additional hardware overhead.

\begin{figure}[t]
\centering
\includegraphics[width=0.7\columnwidth]{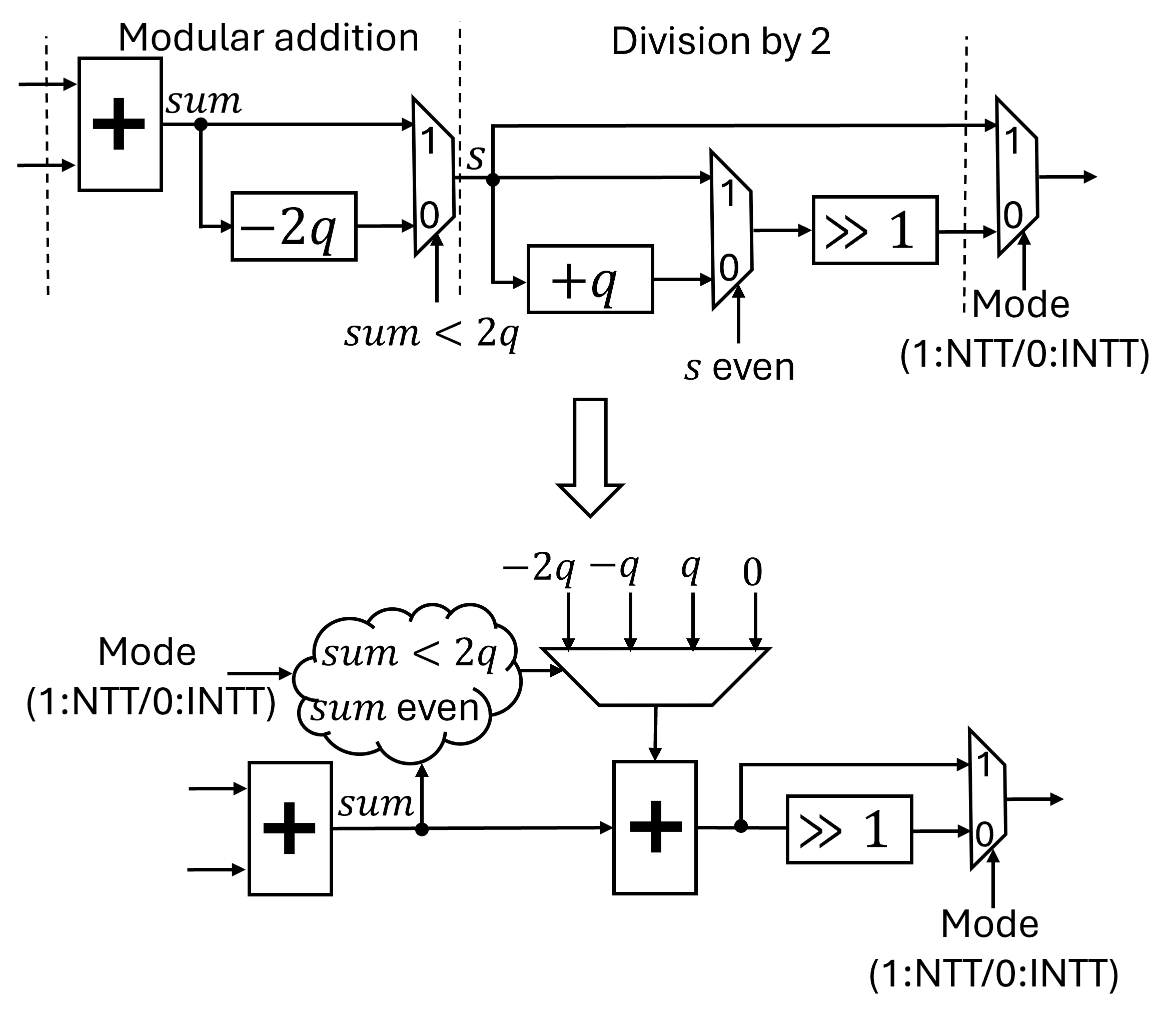}
\caption{The cascaded structure of modular addition, used in both NTT and INTT modes, and scaling by two, which is activated only in INTT mode. Proper range and parity analysis simplifies the serial computation and enables the two operations to be merged in a unified structure.}
\label{fig:adddiv}
\end{figure}

The second division-by-two operation, located at the output of the adder, is optimized by merging its functionality with the adder's modulo correction step \cnum{3}. To illustrate how this is achieved, it is helpful to examine the combined logic of the addition and scaling operations. Fig.~\ref{fig:adddiv} depicts the original cascaded organization of the adder, its modulo correction stage, and the ``divide-by-2'' scaling unit.

The adder internally generates two potential results: $sum$ and $sum-2q$. This ensures that at least one candidate falls within the valid range $[0, 2q-1]$, which is then selected as the output. Those results are needed both in NTT and INTT modes. The subsequent division-by-two stage that is examined only in INTT mode internally processes two candidates: the input and input $+q$. Since $q$ is prime and therefore an odd number, it is guaranteed that at least one of these values will be even. The even one is selected at the output, which can safely be divided by 2 in the modular field~\cite{zhang2020highly}.

Focusing on INTT mode of operation, it is worth noting that the parity of the input to the scaling unit, denoted by \(s\) in Fig.~\ref{fig:adddiv}, is determined directly by the parity of the adder output \(sum\). The value of \(s\) is either \(sum\) or \(sum - 2q\). Since \(q\) is prime, \(2q\) is even, and subtracting an even number does not affect parity. Therefore, in both cases, \(s\) has the same parity as \(sum\). This implies that the combined output of the modulo addition and the ``divide-by-2'' scaling can be selected using the parity of \(sum\) and its range, without requiring these operations to be performed serially. For the parity and the range of $sum$ in INTT mode we can distinguish four cases:
\begin{itemize}
\item 
{\bf $\mathbf{sum < 2q}$ and even:} Since $sum$ is even, it can be scaled directly by 2 (i.e., right-shifted by one position).

\item 
{\bf $\mathbf{sum \ge 2q}$ and even:} Since $sum$ is even, it can be directly scaled by 2. Division by 2 also corrects the range overflow, returning the result to the extended dynamic range $(0, 2q)$. In this case, subtracting $2q$, as done in the original configuration, is redundant.

\item 
{\bf $\mathbf{sum < 2q}$ and odd:} Since $sum$ is odd, it cannot be divided by 2 directly. It must first be increased by $q$ and then scaled accordingly. Since $q$ is prime and thus odd, $sum+q$ is even in this case. Although $sum + q$ may exceed the desired dynamic range, $(sum + q)/2$ always lies within the appropriate range, and no additional correction is required.

\item 
{\bf $\mathbf{sum \ge 2q}$ and odd:}
Since $sum$ is odd, it must first be transformed into an equivalent even integer before scaling. This can be achieved either by increasing $sum$ by $q$, as in the previous case, or by reducing it by $q$. Here, we choose to compute $sum - q$, which is even and after scaling, i.e., $(sum - q)/2$, the result is also brought back into the appropriate extended dynamic range without requiring any additional correction.

\end{itemize}

\begin{figure}[t]
\centering
\includegraphics[width=\columnwidth]{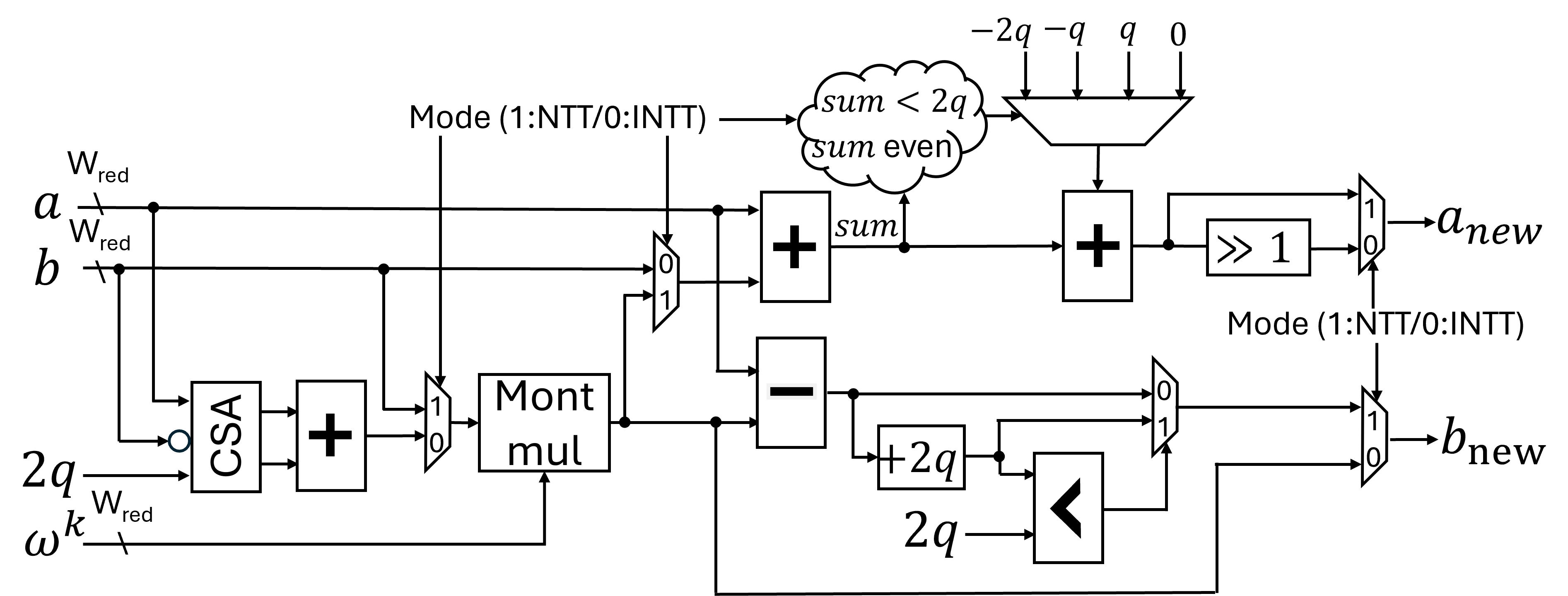}
\caption{The proposed unified butterfly unit supports both NTT and INTT computations. The arithmetic operators represent typical integer arithmetic without modular reduction (only correction step \cnum{4} shown in Fig.~\ref{fig:pe} is kept and integrated in the main datapath). Multiplication is implemented following Montgomery multiplication (Mont. Mul.). Due to extended redundant data representation the bit width of all input, output, and internal signals is \(\text{w}_\text{red} = \left\lceil \log q \right\rceil + 3\) bits. }
\label{fig:pe_div}
\end{figure}

The new block, shown at the bottom of Fig.~\ref{fig:adddiv}, provides a unified implementation of modular addition in NTT mode and modular addition combined with division-by-2 in INTT mode. When \(sum\) is passed to the output unchanged, this is treated as an addition of zero. The selection among the remaining constants, \(+q\), \(-q\), and \(-2q\), is determined by the mode of operation as well as the range and parity of \(sum\). Notably, this solution is generic and orthogonal to the choice of number representation, and it operates correctly even with a canonical range without any redundancy.

The final proposed unified butterfly unit structure that includes all the aforementioned microarchitectural optimizations is shown in Fig.~\ref{fig:pe_div}, that has incorporated the merged structure shown in Fig.~\ref{fig:adddiv}. Multiplication is implemented with a Montgomery multiplier, while add and subtract operators represent typical integer arithmetic without modular reduction. The only modulo correction step that remained refers to the subtraction in NTT mode (correction step \cnum{4} shown in Fig.~\ref{fig:pe}).

\subsection{Mapping Montgomery Multipliers to FPGA DSP blocks}
\label{ss:dsp}

The product of two integers $a, b$ that follow the Montgomery form is given by 
\begin{equation}
m = \frac{ab+q[ab\mu]_R}{R}, \quad \text{where} \quad \mu=-q^{-1}\pmod{R} 
\end{equation}
At the algorithmic level, this operation consists of two full-width multiplications, one multiplication reduced modulo $R$, and a final addition. Since $R$ is a power of two the division by $R$ is trivial in hardware, as it corresponds to discarding the low-order bits of the result. 

The efficiency with which a Montgomery multiplier maps to FPGA logic, and in particular to the built-in DSP blocks, depends largely on how well the bit widths of the integer multipliers match the dimensions of the DSP blocks. DSP48E1 slices support signed $25 \times 18$-bit multiplication with an optional $48$-bit accumulation stage. In the examined PQC applications, the required bit width is determined by the modulus $q$ and the chosen level of redundancy. Table~\ref{tab:logq} highlights the moduli used in common PQC applications and compares the bit widths required for a canonical (non-redundant) representation with those needed in the proposed approach, which assumes a Montgomery radix $R > 8q$.

\begin{table}[t]
    \centering
    \caption{Modulo $q$ and the corresponding bit widths needed for different PQC schemes with canonical ($\log q$) and the proposed redundant representation ($\log q$ + 3).}
    \begin{tabular}{|c|c|c|c|} \hline
        scheme & $q$ & $\lceil\log q\rceil$ & $\lceil\log q\rceil$ + 3 \\ \hline
        ML-KEM~\cite{avanzi2019crystals} & 3329 & 12 & 15 \\
        ML-DSA~\cite{ducas2018crystals} & 8380417 & 23 & 26   \\ 
        FN-DSA~\cite{fouque2018falcon} & 12289 & 14 & 17 \\
        NewHope~\cite{alkimnewhope} & 12289 & 14 &  17 \\\hline
    \end{tabular}
    \label{tab:logq}
\end{table}

\begin{figure}[t]
\centering
\includegraphics[width=0.8\columnwidth]{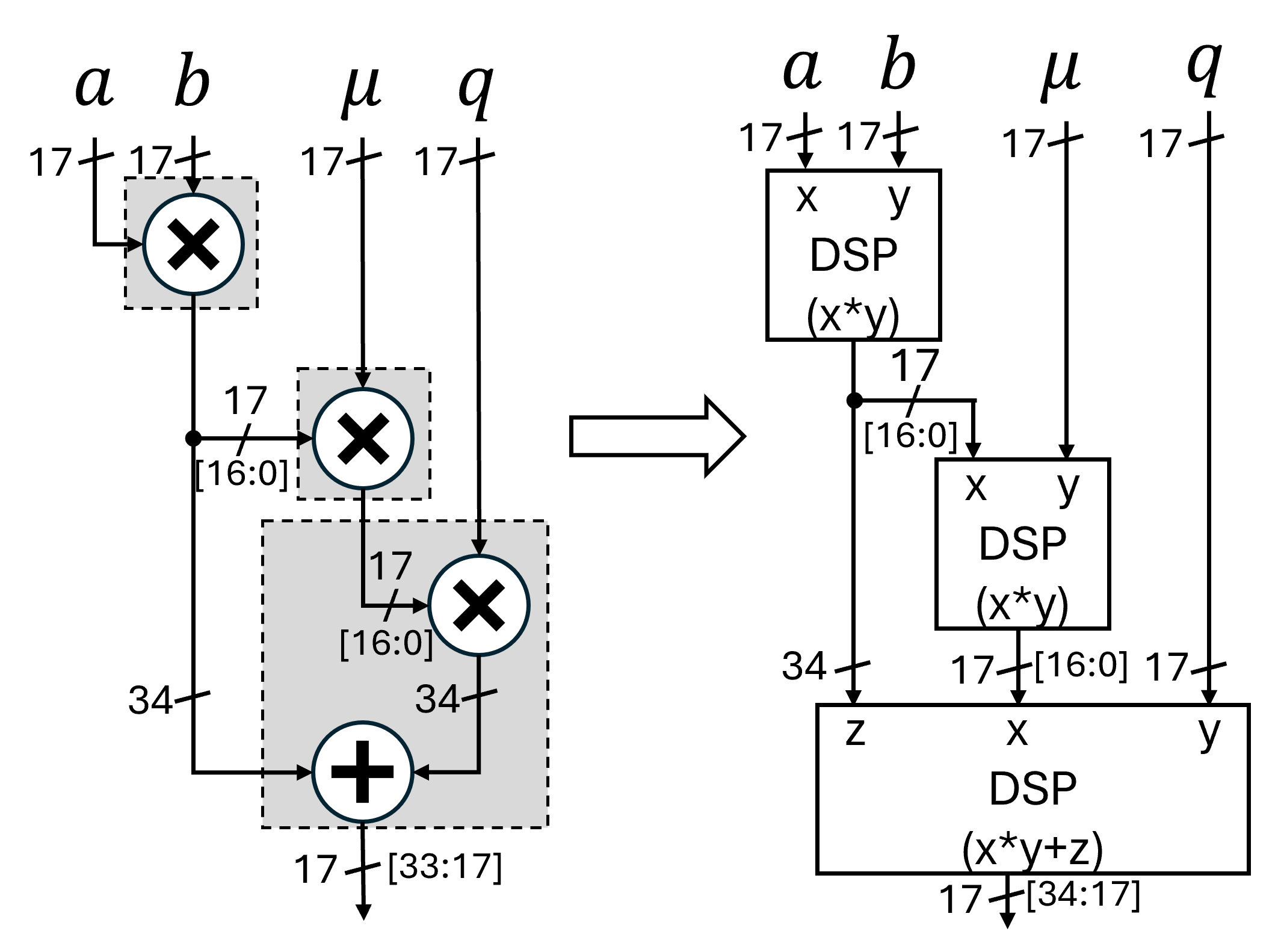}
\caption{Mapping Montgomery multiplication to DSP blocks for inputs of 17 bits as needed by most PQC applications.}
\label{fig:dspmap}
\end{figure}

The bitwidths depicted in Table~\ref{tab:logq} for the most common PQC algorithms belong to two distinct categories with respect to the ease of mapping Montgomery multiplication to DSP blocks. In the first category that includes `ML-KEM', `FN-DSA' (formerly known as Falcon) and `NewHope' the bitwidth needed, even under the introduced redundancy, is less or equal to 17 bits. In this case, each sub-multiplier of Montgomery multiplication accepts 17 bits as input and it can be directly mapped to one DSP block. This is illustrated in Fig.~\ref{fig:dspmap}, where a complete $17$-bit Montgomery multiplier is mapped to 3 DSP blocks. The first two operate as simple integer multipliers while the third implements a combined multiply-add operation.

\begin{figure}[t]
\centering
\includegraphics[width=\columnwidth]{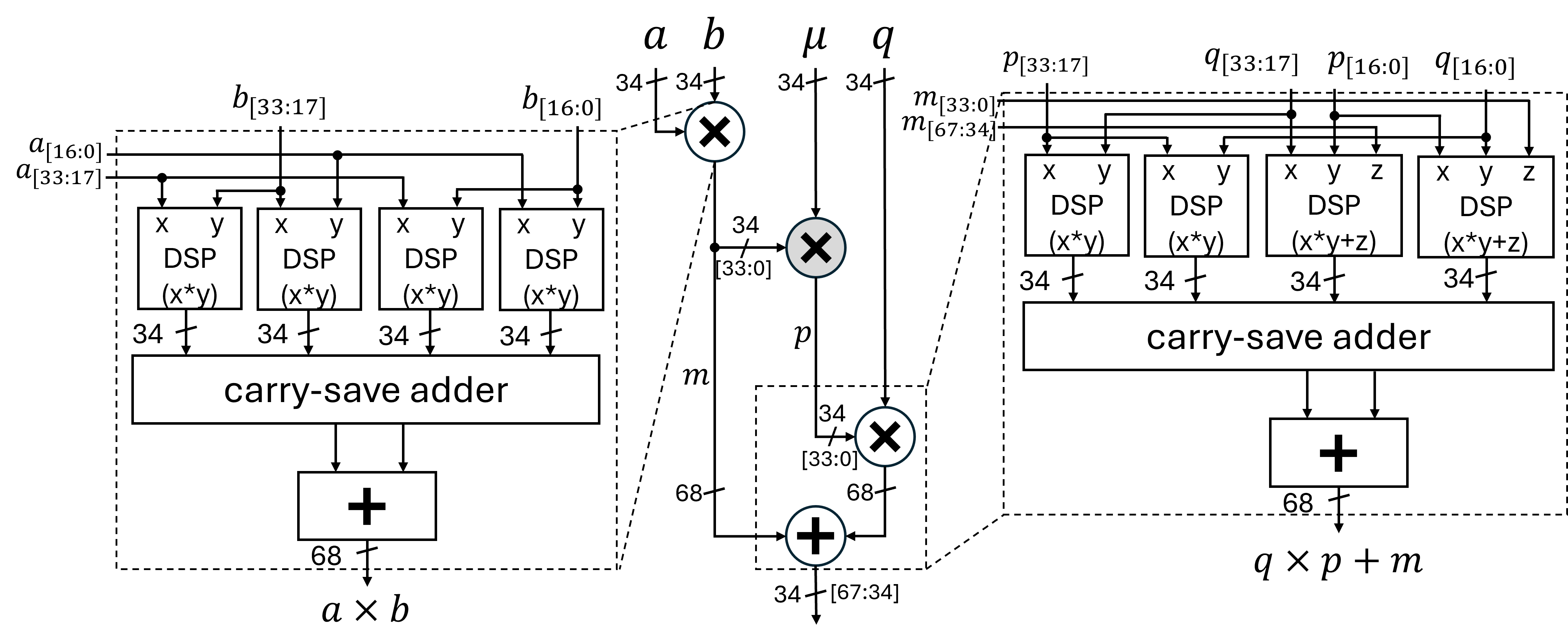}
\caption{Mapping Montgomery multiplication to DSP blocks for 34-bit inputs to support all PQC applications. The 34-bit multiplication and multiply–add blocks that form the Montgomery multiplier are constructed hierarchically from smaller multipliers, each of which maps seamlessly onto existing DSP blocks. The multiplier marked in grey consists of one less DSP block since it computes a result modulo the Montgomery radix $R$ that is power of two.}
\label{fig:big-mult}
\end{figure}
For larger bit widths, as required by ML-DSA, each integer multiplier is implemented hierarchically using multiple DSP blocks. The hierarchical mapping of the two integer multipliers and the multiply–add stage of the Montgomery multiplier is illustrated in Fig.~\ref{fig:big-mult}. In this configuration, the inputs are assumed to be 34 bits wide, and each wide multiplication is decomposed into four 17-bit sub-multipliers, each mapped to a single DSP block. The partial products generated by these sub-multipliers are properly aligned and combined using a carry-save adder, followed by a final carry-propagate adder implemented outside the DSP blocks.

The second wide multiplication, involving the Montgomery inverse $\mu$, is performed modulo $R$. As a result, only three of the four DSP blocks are required in this case, since the most significant sub-multiplier does not contribute to the final result. Accordingly, both this DSP block and its associated carry-save logic are omitted.

For the final multiply–add stage of the Montgomery multiplier, carry-save addition is avoided altogether. Instead, partial products are accumulated using the adders available within the DSP blocks. Specifically, two of the four DSP blocks are configured in multiply–add mode, allowing the addition to be integrated directly into the partial-product accumulation. This approach eliminates the need for a separate carry-save stage and further reduces overall resource utilization.

Overall, a 34-bit Montgomery multiplier that covers the case of moduli up to 31 bits (due to redundant number representation) is built using 11 DSP blocks in total.

Finally, to meet the latency requirements imposed by the non-conflict, bubble-free architecture in \cite{mu2023scalable} adopted by the proposed architecture, each DSP block is configured to operate with two pipeline stages. As a result, the Montgomery multiplier employs six pipeline stages in total, leaving two additional stages available for other PE operations. Although this choice may limit the maximum achievable clock frequency, it enables a more scalable and configurable PE array while preserving a DSP-centric computation model.
\section{Experimental Results}
\label{s:exp}

The goal of this section is to evaluate the hardware cost and performance impact of the proposed redundant Montgomery-based NTT accelerator. We first focus on evaluating the efficiency of the proposed butterfly functional unit (BFU) against prior designs. We then focus on how these BFU-level improvements translate to full NTT accelerator performance. All designs were synthesized using Vivado 2025.2. Unless stated otherwise, the Virtex-7 FPGA is used to enable direct comparison with prior work.

\subsection{Comparison of Unified Butterfly Units}
The BFU is the key building block that differentiates NTT architectures. Several prior works report detailed per-BFU results, enabling a direct, apples-to-apples comparison with our design. Table~\ref{tab:bfu} summarizes the relevant metrics that characterize the performance of each BFU when mapped to a specific FPGA platform. The second column lists the target FPGA platform for each architecture cited in the first column, while the third column specifies the modular reduction technique employed in each case, with Montgomery-based variants used in the majority of designs. The fourth column reports the bit width considered for each BFU targeting PQC applications. The remaining columns present the hardware characteristics of each BFU, including the configured latency (in cycles), the achieved clock frequency, and FPGA resource utilization in terms of LUTs, FFs, and DSP blocks.

\begin{table}[t]
    \centering
    \caption{Comparison of the proposed unified BFU with state-of-the-art}
    \begin{adjustbox}{center}    
    \begin{tabular}{|c|cccccccc|}\hline
    work & board & reduction & \makecell[c]{$\lceil \log q\rceil$ \\ (bits)} & \makecell[c]{clock \\ cycles} & LUT & FF & DSP & \makecell[c]{freq \\ (MHz)} \\ \hhline{=========}
     \cite{krieger2025openntt} & V7 & WLM & 30 & 14 & 380 & 449 & 7 & 290 \\\hhline{=========}
     \cite{yang2023hardware} & A7 & M* & 12 & 4 & 160 & 109 & 1 & - \\\hhline{=========}
     \cite{Yang2022nttgen} & U200 & B & 30 & 14 & 1081 & 920 & 12 & 400 \\\hhline{=========}
     \multirow{2}{*}{\cite{kemal2022cohantt}} & V7 & WLM & 12-30 & 4-7 & 703 & 474 & 8 & 174 \\ \hhline{~--------}
     & A7 & WLM & 12-30 & 4-7 & 705 & 488 & 8 & 117 \\ \hhline{=========}
     \multirow{3}{*}{\cite{mu2023scalable}} & \multirow{3}{*}{V7} & M & 13 & 7& 173 & 122 & 3 & 286 \\
     & & M & 14 & 7 & 132 & 131 & 3 & 278 \\
     & & M & 24 & 7 & 354 & 271 & 4 & 185 \\\hhline{=========}
     \multirow{3}{*}{\cite{kumarathunga2025autontt}} & \multirow{3}{*}{U200} & B & 30 & 10 & 766 & 403 & 11 & 250 \\
     & & M & 30 & 10 & 813 & 434 & 11 & 250 \\
     & & WLM & 30 & 10 & 635 & 429 & 10 & 250 \\\hhline{=========}
     \multirow{6}{*}{\makecell[c]{This \\ work}} & \multirow{2}{*}{V7} & M-red & up to 14 & 8 & 209 & 194 & 3 & 437 \\
     & & M-red & up to 31 & 8 & 580 & 481 & 11 & 391 \\ \hhline{~--------}
     & \multirow{2}{*}{A7} & M-red & up to 14 & 8 & 225 & 162 & 3 & 358 \\
     & & M-red & up to 31 & 8 & 580 & 481 & 11 & 243 \\ \hhline{~--------}
     & \multirow{2}{*}{U200} & M-red & up to 14 & 8 & 225 & 162 & 3 & 905 \\
     & & M-red & up to 31 & 8 & 572 & 501 & 11 & 496 \\ \hline
    \end{tabular}
    \end{adjustbox}
    M: Montgomery. WLM: Word Level Montgomery. B: Barrett. M*: Modified Montgomery. M-red: Redundant Montgomery.
    \label{tab:bfu}
\end{table}

For the proposed design we implemented two variants for BFUs: (a) 17-bit BFUs supporting moduli up to 14 bits and (b) 34-bit BFUs that work for moduli up to 31 bits.

The key result from Table~\ref{tab:bfu} is the increased operating frequency achieved by the proposed BFUs across all evaluated FPGA platforms\footnote{each frequency comparison is made within the same board.}.
On Virtex-7, the 17-bit BFU reaches 437MHz, compared to 286MHz for the Montgomery BFU in~\cite{mu2023scalable} at a similar bit-width, corresponding to a 1.53$\times$ frequency improvement. When compared to Coha-NTT~\cite{kemal2022cohantt}, which operates at 174MHz on Virtex-7, the frequency gain increases to 2.5$\times$. For wider datapaths, the proposed 34-bit BFU operates at 391MHz on Virtex-7, exceeding the 185MHz reported in~\cite{mu2023scalable} for 24-bit arithmetic by 2.1$\times$, while also supporting larger moduli (up to 31 bits). Similar trends are observed on the Alveo U200 platform. The proposed 17-bit BFU achieves 905MHz\footnote{Maximum achievable post-implementation frequency for an isolated BFU.}, compared to 400MHz in~\cite{Yang2022nttgen} and 250MHz in~\cite{kumarathunga2025autontt}, corresponding to 2.26$\times$ and 3.6$\times$ higher frequency, respectively.
This improvement stems from two design choices:
\begin{itemize}
    \item Montgomery multiplication is fully absorbed into DSP datapaths with minimal LUT and FF overhead, and
    \item the proposed redundant representation removes intermediate modular reductions, shortening the critical adder and subtractor paths.
\end{itemize}

In terms of resource usage, the proposed 34-bit BFU is more efficient than other wide-precision designs targeting similar bit-widths, despite supporting runtime-configurable moduli. For smaller bit-widths, some prior works report lower LUT usage; however, these designs either hard-code the modulus or rely on specialized arithmetic optimizations that limit generality. 

In contrast, the proposed BFUs support a programmable modulus that can be changed at runtime while achieving substantially higher operating frequencies. When the modulus is changed, the table containing the new twiddle factors, which are computed offline, is loaded into the design. In this way, no additional on chip computation is required for twiddle factor generation, and the datapath remains unchanged. The supported modulus range is determined solely by the datapath width of the BFU. Apart from the standard mathematical requirements of the NTT that the modulus is an odd prime admitting the required roots of unity, that is, $q \equiv 1 \pmod{2N}$, the architecture imposes no additional constraints. Consequently, the design is applicable not only to the Kyber and Dilithium parameter sets, but also to any NTT friendly prime within the supported bit width range.


\subsection{Comparison of complete NTT Accelerators}
The efficient microarchitecture of the proposed unified butterfly unit leads to more efficient NTT accelerators. To quantify the benefits of the proposed architecture, we compare it against three state-of-the-art approaches.

Specifically, we compare against~\cite{kemal2022cohantt}, which emphasizes run-time configurability of the operating modulus. This is directly relevant to our design, as the modulus in our accelerator is also run-time configurable, whereas most prior works generate separate designs tailored to a single PQC scheme. In addition, we include the NTT accelerators of~\cite{mert2022an}, which employ algorithmic optimizations such as word-level Montgomery arithmetic, along with efficient hardware mapping through extensive DSP utilization. Finally, we compare against the design of~\cite{mu2023scalable}, which is closely related to the proposed approach, as both follow the same memory architecture and data access pattern. The primary difference lies in the microarchitecture of the processing elements, allowing us to highlight the overall impact of the proposed enhancements at the full NTT accelerator level.

To ensure fair and straightforward conclusions, all comparisons are performed on the same Virtex-7 FPGA platform. Table~\ref{tab:ntt} summarizes the performance results for all evaluated configurations. We assess NTT sizes of 256, 512, and 1024 using \(1\times 1\), \(w\times 1\), and \(w\times 2\) PE arrays, where applicable. These configurations operate across a range of data bit widths considered in prior state-of-the-art designs and collectively cover all relevant PQC applications. 

\begin{table}[H]
\renewcommand{\arraystretch}{0.9}
    \centering
    \caption{Comparison of the proposed NTT accelerator with state-of-the-art}
    \begin{adjustbox}{center}    
    \begin{tabular}{|c|ccc|cccc|cccc|}\hline
    work & \makecell[c]{$\lceil \log q\rceil$ \\ (bits)} & $N$ & PEs & LUT & FF & DSP & BRAM & \makecell[c]{clock \\ cycles} & \makecell[c]{freq \\ (MHz)} & \makecell[c]{time \\ ($\mu$s)} & {\makecell[c]{power \\ (W)}} \\ \hhline{============}
     \multirow{3}{*}{\cite{kemal2022cohantt}} & \multirow{3}{*}{12-30} & \multirow{3}{*}{256} & 1 & 2128 & 1144 & 8 & 3 & 1052 & 174 & 6.0 & {-} \\ 
     & & & 8 & 11000 & 5422 & 64 & 12 & 138 & 186 & 0.7 & {-} \\ 
     & & & 32 & 61000 & 17000 & 64 & 12 & 84 & 167 & 0.5 & {-} \\ \hhline{============}
     \multirow{8}{*}{\cite{mert2022an}} & \multirow{2}{*}{13} & \multirow{2}{*}{256} & 1 & 489 & - & 3 & 2.5 & 1056 & 125 & 8.4 & {-} \\
     & & & 8 & 2371 & - & 24 & 12 & 160 & 125 & 1.3 & {-} \\ \hhline{~-----------}
     & \multirow{2}{*}{14} & \multirow{2}{*}{1024} & 1 & 575 & - & 3 & 11 & 5160 & 125 & 41.5 & {-} \\
     & & & 32 & 17188 & - & 96 & 48 & 200 & 125 & 1.6 & {-} \\ \hhline{~-----------}
     & \multirow{2}{*}{23} & \multirow{2}{*}{256} & 1 & 888 & - & 7 & 5 & 1096 & 125 & 8.8 & {-} \\
     & & & 8 & 5071 & - & 56 & 12 & 200 & 125 & 1.6 & {-} \\ \hhline{~-----------}
     & \multirow{2}{*}{29} & \multirow{2}{*}{1024} & 1 & 966 & - & 7 & 21.5 & 5210 & 125 & 41.7 & {-} \\
     & & & 32 & 38072 & - & 224 & 48 & 250 & 125 & 2.0 & {-} \\\hhline{============}
     \multirow{8}{*}{\cite{mu2023scalable}} & \multirow{2}{*}{13} & \multirow{2}{*}{256} & 1 & 449 & 271 & 3 & 3 & 1031 & 286 & 3.6 & {-} \\
     & & & 8 & 6245 & 1864 & 24 & 12 & 135 & 256 & 0.5 & {-} \\ \hhline{~-----------}
     & \multirow{4}{*}{14} & \multirow{2}{*}{512} & 1 & 489 & 245 & 3 & 3 & 2311 & 278 & 8.3 & {-} \\
     & & & 16 & 28616 & 4211 & 48 & 24 & 151 & 154 & 1.0 & {-} \\ \hhline{~~----------}
     & & \multirow{2}{*}{1024} & 1 & 515 & 306 & 3 & 3 & 5127 & 278 & 18.4 & {-} \\
     & & & 8$\times$2 & 8515 & 3618 & 49 & 12 & 334 & 172 & 1.9 & {-} \\ \hhline{~-----------}
     & \multirow{2}{*}{24} & \multirow{2}{*}{256} & 1 & 691 & 451 & 5 & 3 & 1031 & 187 & 5.5 & {-} \\
     & & & 2$\times$2 & 2466 & 1637 & 20 & 4.5 & 270 & 175 & 1.5 & {-} \\\hhline{============}
     \multirow{18}{*}{\makecell[c]{This \\ work}} & \multirow{9}{*}{\makecell[c]{14 \\ (14$+$3)}} & \multirow{3}{*}{256} & 1 & 485 & 343 & 3 & 3 & 1032 & 437 & 2.3 & {0.78} \\
     & & & 8 & 6608 & 2194 & 24 & 12 & 136 & 402 & 0.3 & {1.87} \\ 
     &  &  & 2$\times$2 & 1546 & 1164 & 12 & 5 & 272 & 424 & 0.6 & {1.08} \\ \hhline{~~----------}
     &  & \multirow{3}{*}{512} & 1 & 512 & 351 & 3 & 3 & 2312 & 425 & 5.4 & {0.78} \\
     &  &  & 16 & 29221 & 5168 & 48 & 24 & 152 & 258 & 0.5 & {3.25} \\ 
     &  &  & 4$\times$2 & 3584 & 1992 & 24 & 8 & 304 & 280 & 1.0 & {1.22} \\ \hhline{~~----------}
     &  & \multirow{3}{*}{1024} & 1 & 592 & 369 & 3 & 3 & 5128 & 424 & 12.1 & {0.79} \\ 
     &  &  & 32 & 139172 & 12144 & 96 & 48 & 168 & 210 & 0.8 & {9.0} \\ 
     &  &  & 8$\times$2 & 9942 & 5131 & 48 & 12 & 336 & 270 & 1.2 & {1.97} \\ \hhline{~-----------}
     & \multirow{9}{*}{\makecell[c]{31 \\ (31$+$3)}} & \multirow{3}{*}{256} & 1 & 951 & 683 & 11 & 3 & 1032 & 390 & 2.6 & {0.90} \\ 
     &  &  & 8 & 9642 & 4709 & 88 & 12 & 136 & 358 & 0.4 & {2.70} \\
     &  &  & 2$\times$2 & 3616 & 2108 & 44 & 5 & 272 & 380 & 0.7 & {1.57} \\ \hhline{~~----------}
     &  & \multirow{3}{*}{512} & 1 & 1104 & 768 & 11 & 3 & 2312 & 382 & 6.0 & {0.92} \\
     &  &  & 16 & 36618 & 10276 & 176 & 24 & 152 & 238 & 0.6 & {4.42} \\ 
     &  &  & 4$\times$2 & 6733 & 4487 & 88 & 8 & 304 & 255 & 1.2 & {1.86} \\ \hhline{~~----------}
     &  & \multirow{3}{*}{1024} & 1 & 1276 & 808 & 11 & 3 & 5128 & 380 & 13.5 & {0.93} \\
     &  &  & 32 & 158141 & 14289 & 352 & 48 & 168 & 200 & 0.9 & {10.8} \\
     &  &  & 8$\times$2 & 15390 & 10236 & 176 & 12 & 336 & 245 & 1.4 & {3.14} \\\hline
    \end{tabular}
    \end{adjustbox}
    \label{tab:ntt}
\end{table}

Table~\ref{tab:ntt} is structured into three column groups. The first group describes the accelerator configuration parameters. The second group details FPGA resource utilization, including LUTs, FFs, DSP blocks, and BRAMs. The third group reports performance metrics, comprising the achieved clock frequency, the total number of clock cycles required to complete the computation, the corresponding execution time in absolute terms, obtained by multiplying the clock cycle count by the clock period for each configuration, and power consumption. Related works used in the comparisons do not report power measurements, preventing direct power comparisons.

First, we analyze the behavior of scalar configurations (PEs $=$ 1$\times$1) as the NTT size $N$ and modulus bit-width $q$ increase. Across all architectures, increasing $N$ from 256 to 1024 leads to an almost linear growth in clock cycles (e.g., from $~$1030 to $~$5130 cycles for $\lceil\log q\rceil=$ 14 in the proposed work), reflecting the inherent $O(N\log N)$ complexity of the NTT. Similarly, increasing the modulus bit-width results in higher LUT, FF, and DSP utilization due to wider arithmetic datapaths, while the cycle count remains largely unaffected. This trend is consistently observed in Table~\ref{tab:ntt} for~\cite{kemal2022cohantt},~\cite{mert2022an},~\cite{mu2023scalable}, and this work. Focusing on identical operating points (e.g., $N=$ 1024, $\lceil\log q\rceil=$ 14, PEs $=$ 1$\times$1), our design achieves comparable cycle counts (5128 vs. 5127 cycles in~\cite{mu2023scalable}) but operates at significantly higher frequency (424MHz vs. 278MHz), resulting in a 34\% reduction in execution time (12.1$\mu$s vs. 18.4$\mu$s). Compared to~\cite{mert2022an}, whose frequency is fixed at 125MHz, the runtime is reduced by 71\% for the same configuration, clearly indicating that frequency scaling dominates performance when utilizing only one PE.

We next examine the effect of increasing parallelism considering the case of one-dimensional PE arrays ($w\times$1). As expected, increasing $w$ significantly reduces the number of clock cycles, at the cost of higher hardware utilization. For instance, for $N=$ 1024 and $\lceil\log q\rceil=$ 14, moving from 1 PE to 32$\times$1 PEs reduces the cycle count from 5128 to 168 in the proposed design, while LUT usage increases significantly (from 592 to 139k LUTs). This tradeoff is consistently observed across all designs under comparison. 

However, important differences emerge when comparing the scaling of clock frequency and overall runtime. In~\cite{kemal2022cohantt}, aggressive parallelization leads to diminishing frequency returns (e.g., 167MHz at $N=$ 256, PEs $=$ 32 and a maximum $\lceil\log q\rceil$ of 30). Of course, the proposed design is also affected by this issue; nevertheless, it maintains a relatively high operating frequency (200MHz at PEs $=$ 32 for $N=$ 1024 and $\lceil\log q\rceil=$ 31), yielding shorter execution times despite similar cycle counts. Compared to~\cite{mert2022an}, which reports PE $=$ 32 designs at a fixed 125MHz, the proposed accelerator achieves up to 55\% lower runtime  for equivalent $N$ and $w$ (0.9$\mu$s vs. 2.0$\mu$s for $\lceil\log q\rceil=$ 31 and 29 respectively, $N=$ 1024 and $w=$ 32), even though the design supports runtime-configurable moduli and therefore incurs higher area. These results highlight that while increasing $w$ is effective in reducing latency, frequency scalability becomes a critical differentiating factor across architectures.

Finally, we focus on two-dimensional PE organizations ($w\times$2), which consistently emerge as a balanced design point between performance and hardware cost. Compared to their $w\times$1 counterparts, 2D PE arrays substantially reduce LUT, DSP, and BRAM usage while preserving a large fraction of the throughput gains. For example, for 
$N=$ 1024 and $\lceil\log q\rceil=$ 14, the PEs $=$ 8$\times$2 configuration of the proposed work uses 14$\times$ fewer LUTs than the PEs $=$ 32$\times$1 configuration (9.9k vs. 139k LUTs), while achieving an execution time of 1.2$\mu$s compared to 0.8$\mu$s—trading 33\% lower execution time for an order-of-magnitude area reduction. Similar trends are observed for $N=$ 256 and $N=$ 512, and across both supported modulus widths. When compared against~\cite{mu2023scalable}, which also evaluates 2D-like configurations, the proposed design consistently achieves higher clock frequencies (e.g. 270MHz vs. 172MHz for $N=$ 1024, $\lceil\log q\rceil=$ 14 and PEs $=$ 8$\times$2), resulting in 36\% lower execution time at comparable cycle counts. 
Overall, these results demonstrate that 2D PE arrays provide a highly attractive operating point, delivering a near-optimal trade-off among throughput, resource utilization, and power consumption, and therefore constituting the most practical choice across a broad range of architectures and parameter sets.

\begin{table}[th]
    
\renewcommand{\arraystretch}{0.9}
    \centering
    \caption{Comparison of the proposed NTT accelerator with the BRAM-free architecture~\cite{ni2023towards} for 12-bit operands and polynomial degree 256. }
    \begin{adjustbox}{center}    
    \begin{tabular}{|c|c|cccc|ccc|}\hline
    work  & PEs & LUT & FF & DSP & BRAM & \makecell[c]{clock \\ cycles} & \makecell[c]{freq \\ (MHz)} & \makecell[c]{time \\ ($\mu$s)}  \\ \hhline{=========}
     \cite{ni2023towards} &  2 & 1154 & 1031 & 2 & 0 & 456 & 300 & 1.5 \\\hhline{=========}
     \multirow{3}{*}{\makecell[c]{This \\ work}} & 1 & 485 & 343 & 3 & 3 & 1032 & 437 & 2.3 \\
      & 8 & 6608 & 2194 & 24 & 12 & 136 & 402 & 0.3 \\ 
      & 2$\times$2 & 1546 & 1164 & 12 & 5 & 272 & 424 & 0.6 \\ \hhline{---------}
    \end{tabular}
    \end{adjustbox}
    \footnotesize
    \label{tab:bramfree}
\end{table}

Additionally, Table~\ref{tab:bramfree} highlights the trade-off between area cost and performance when comparing the proposed architecture with the BRAM-free design of~\cite{ni2023towards}, to provide a more complete comparison against works that targets memory optimizations. The BRAM-free implementation achieves low latency using two processing elements and no block memories, but at the cost of significantly higher LUT and FF utilization. In contrast, our single-PE configuration demonstrates a substantially lower area footprint, reducing LUT usage by more than 2$\times$ while still operating at a higher clock frequency, albeit with increased latency. At the same time, the scalability of the proposed architecture allows multiple design points that balance cost and performance. For example, the 2$\times$2 and 8-PE configurations provide progressively higher throughput and significantly lower execution time, reaching up to a 5$\times$ speedup compared to~\cite{ni2023towards}. Overall, the results demonstrate that the proposed architecture spans a wide design space, offering both low-cost and high-performance implementations depending on system requirements.
\section{Conclusions}
\label{s:con}

This work presented a high-performance FPGA-oriented NTT accelerator designed to meet the demanding requirements of post-quantum cryptography. The proposed architecture combines arithmetic-level innovation with targeted microarchitectural optimizations to reduce latency, improve frequency scalability, and maintain flexibility for runtime-programmable moduli. 

First, we introduced a unified redundant Montgomery-based number representation that supports both Montgomery multiplication and combined subtract-multiply operations without requiring conditional modulo corrections. By extending Montgomery redundancy to cover the arithmetic patterns of both NTT and INTT butterflies, the proposed representation enables a single unified butterfly unit to operate efficiently in both modes while maintaining correctness. 
Second, we proposed a hardware-efficient treatment of INTT scaling operations. One division-by-two operation is fully removed by shifting its cost to offline twiddle-factor precomputation, while the remaining scaling operation is merged with the modular addition logic through careful range and parity analysis. This approach removes dedicated scaling units, shortens the critical path and improves the achievable clock frequency. Third, we developed a DSP-centric Montgomery multiplier mapping strategy that hierarchically decomposes wide multiplications into sub-operations that align naturally with FPGA DSP block dimensions. 

Together, these contributions enable NTT accelerators that achieve consistently higher clock frequencies without increasing pipeline depth, leading to by 35\% to 73\% reductions in absolute execution time across a wide range of NTT sizes, degrees of parallelism, and runtime-configurable moduli. Experimental results demonstrate that the proposed architecture outperforms state-of-the-art configurable NTT accelerators in latency while maintaining comparable or lower hardware cost, and remains competitive even when compared against fixed-modulus designs. 

Overall, this work shows that careful co-design of arithmetic representations and microarchitecture is a powerful approach for advancing scalable, high-performance NTT hardware accelerators required for future post-quantum cryptographic systems. 
However, computational efficiency addresses only part of the challenge. Securing these accelerators against physical vulnerabilities is a mandatory requirement for real-world deployment. Therefore, our future work will focus on integrating robust countermeasures against side-channel attacks, ensuring that the theoretical guarantees of PQC are not undermined by physical hardware security vulnerabilities.



\section*{Acknowledgments}
This work is supported by a research grant of Nokia Bell Labs to Democritus University of Thrace for "Hardware accelerators for Privacy Preserving Technologies"

\bibliographystyle{elsarticle-num} 
\bibliography{refs}
\end{document}